\documentclass[british,english]{article}
\usepackage[T1]{fontenc}
\usepackage[latin9]{inputenc}
\usepackage{geometry}
\usepackage{color}
\usepackage{float}
\usepackage{amsmath}
\usepackage{amssymb}
\usepackage{graphicx}
\usepackage[authoryear]{natbib}

\makeatletter

\providecommand{\tabularnewline}{\\}
\floatstyle{ruled}
\newfloat{algorithm}{tbp}{loa}
\providecommand{\algorithmname}{Algorithm}
\floatname{algorithm}{\protect\algorithmname}

\usepackage{bbm}

\usepackage{algpseudocode}
\usepackage{algorithm}

\@ifundefined{showcaptionsetup}{}{%
 \PassOptionsToPackage{caption=false}{subfig}}
\usepackage{subfig}
\makeatother

\usepackage{babel}
\addto\captionsbritish{\renewcommand{\algorithmname}{Algorithm}}

\begin{document}

\title{Marginal sequential Monte Carlo for doubly intractable models}

\author{Richard G. Everitt, Dennis Prangle, Philip Maybank and Mark Bell}
\maketitle
\begin{abstract}
Bayesian inference for models that have an intractable partition function
is known as a \emph{doubly intractable} problem, where standard Monte
Carlo methods are not applicable. The past decade has seen the development
of auxiliary variable Monte Carlo techniques \citep{Moller2006,Murray2006}
for tackling this problem; these approaches being members of the more
general class of pseudo-marginal, or exact-approximate, Monte Carlo
algorithms \citep{Andrieu2009}, which make use of unbiased estimates
of intractable posteriors. \citet{Everitt2017} investigated the use
of exact-approximate importance sampling (IS) and sequential Monte
Carlo (SMC) in doubly intractable problems, but focussed only on SMC
algorithms that used data-point tempering. This paper describes SMC
samplers that may use alternative sequences of distributions, and
describes ways in which likelihood estimates may be improved adaptively
as the algorithm progresses, building on ideas from \citet{Moores2015}.
This approach is compared with a number of alternative algorithms
for doubly intractable problems, including approximate Bayesian computation
(ABC), which we show is closely related to the method of \citet{Moller2006}.
\end{abstract}

\section{Introduction}

\subsection{Background}

\subsubsection{The pseudo-marginal approach\label{subsec:The-pseudo-marginal-approach}}

Pseudo-marginal, or exact-approximate, Monte Carlo algorithms \citep{Beaumont2003,Andrieu2009,Fearnhead2010a}
are one of the main developments in Bayesian computation in the past
decade. Suppose that we wish to simulate from a posterior distribution
$\pi\left(\theta\mid y\right)$ determined by prior $p\left(\theta\right)$
and likelihood $f\left(y\mid\theta\right)$, where $\theta$ has dimension
$d$. The pseudo-marginal idea permits the construction of a Monte
Carlo method with target $\pi$ in the case where only an unbiased
estimate $\hat{\pi}$ of the density is available at every $\theta$.
In this paper we focus on the case where the density $\pi$ is ``intractable'',
i.e. cannot be evaluated pointwise at $\theta$, due to the likelihood
being intractable.

In this case, suppose that, for any $\theta$, it is possible to compute
an unbiased estimate $\widehat{f}\left(y|\theta\right)$ of $f\left(y|\theta\right)$.
Then
\begin{enumerate}
\item Using the acceptance probability 
\[
\alpha\left(\theta,\theta^{*}\right)=\min\left\{ 1,\frac{\widehat{f}(y|\theta^{*})p(\theta^{*})q(\theta|\theta^{*})}{\widehat{f}(y|\theta)p(\theta)q(\theta^{*}|\theta)}\right\} ,
\]
where $\theta$ is the current state of a Markov chain and $\theta^{*}$
is the proposed state, yields an MCMC algorithm with target distribution
$\pi\left(\theta|y\right)$.
\item Using the (unnormalised) weight
\[
w=\frac{\widehat{f}(y|\theta)p(\theta)}{q(\theta)},
\]
where $\theta$ an importance point simulated from $q$, yields an
importance sampling algorithm with target distribution $\pi\left(\theta|y\right)$.
\end{enumerate}
It is straightforward to see why this is the case by writing the joint
distribution of \emph{all} of the variables that are being used $\widehat{f}(y|\theta,x)p(x|\theta)p(\theta)$,
where $x$ are the random variables used to generate the estimate
$\widehat{f}$. We then see that an algorithm that simulates from
$\pi(\theta,x|y)$ has the correct marginal distribution, and further
that $q\left(\left(\theta^{*},x^{*}\right)|\left(\theta^{(p)},x^{(p)}\right)\right)=q(\theta^{*}|\theta^{(p)})p(x^{*}|\theta^{*})$
as a proposal within a Metropolis-Hastings algorithm yields the desired
acceptance probability (with a similar extended space representation
being used for the importance sampling case).

\subsubsection{Doubly intractable distributions\label{subsec:Doubly-intractable-distributions}}

In this paper we focus on the case where $f\left(y\mid\theta\right)=\gamma\left(y\mid\theta\right)/Z\left(\theta\right)$
cannot be evaluated pointwise due to the presence of an intractable
partition function $Z\left(\theta\right)$. In this case an MCMC algorithm
targeting $\pi$has acceptance probability
\begin{equation}
\alpha\left(\theta,\theta^{*}\right)=\min\left\{ 1,\frac{\gamma(y|\theta^{*})p(\theta^{*})q(\theta|\theta^{*})}{\gamma(y|\theta)p(\theta)q(\theta^{*}|\theta)}\frac{Z\left(\theta\right)}{Z\left(\theta^{*}\right)}\right\} ,\label{eq:mcmc_Z}
\end{equation}
and an importance sampler targeting $\pi$ has weight
\[
w^{(p)}=\frac{\gamma(y|\theta)p(\theta)}{q(\theta)}\frac{1}{Z\left(\theta\right)}.
\]
These ``ideal'' algorithms are intractable due to the presence of
$Z\left(\cdot\right)$. The following methods circumvent evaluating
$Z\left(\cdot\right)$, leading to tractable algorithms.
\begin{enumerate}
\item \citet{Moller2006} constructs a pseudo-marginal MCMC algorithm by
using an unbiased importance sampling estimator
\begin{equation}
\frac{1}{Z\left(\theta\right)}\approx\frac{q_{u}\left(x\mid\theta,y\right)}{\gamma\left(x\mid\theta\right)},\label{eq:sav}
\end{equation}
where $x\sim f\left(\cdot\mid\theta\right)$, in the numerator in
the acceptance probability (with an analogous estimator being used
in the denominator). $q_{x}$ may be any (normalised) distribution,
but in practice is often chosen as $q_{x}\left(x\mid\theta^{*},y\right)=f\left(x\mid\hat{\theta}\right)$,
where $\hat{\theta}$ is a point estimate such as the maximum likelihood
estimate. We refer to this approach as single auxiliary variable (SAV)
MCMC. The estimator in equation \ref{eq:sav} may be improved, and
remains unbiased, when using $M$ importance points, and/or using
annealed importance sampling (AIS) \citep{Neal2001} with $a$ intermediate
targets in place of standard IS. Let
\begin{eqnarray*}
f_{i}(\cdot|\theta,\hat{\theta},y) & \propto & \gamma_{i}(\cdot|\theta,\hat{\theta},y)\\
 & = & \gamma(\cdot|\theta){}^{\left(i-1\right)/\left(a-1\right)}\gamma(\cdot|\hat{\theta}){}^{\left(a-i\right)/\left(a-1\right)}
\end{eqnarray*}
be a sequence of intermediate targets, with $K_{i}$ an MCMC kernel
with target $f_{i}$. Then if, for each $i$, $x_{i}\sim K_{i}\left(\cdot\mid x_{i+1}\right)$,
our improved estimator is
\begin{equation}
\frac{Z\left(\hat{\theta}\right)}{Z(\theta)}\approx\frac{1}{M}\sum_{m=1}^{M}\prod_{i=2}^{a}\frac{\gamma_{i-1}(x_{i}^{(m)}|\theta,\widehat{\theta},y)}{\gamma_{i}(x_{i}^{(m)}|\theta,\widehat{\theta},y)},\label{eq:mav}
\end{equation}
noting that the additional term $Z\left(\hat{\theta}\right)$ cancels
in the acceptance ratio. The use of AIS in place of IS was proposed
by \citet{Murray2006}, where the approach is referred to as the ``multiple
auxiliary variable'' (MAV) method. \citet{Everitt2017} uses the
estimator in equation \ref{eq:mav} within IS, in which we see that
the term $Z\left(\hat{\theta}\right)$ is the same for every importance
point, thus cancels when using normalised IS estimates.
\item \citet{Murray2006} uses an exact-approximate MCMC algorithm by using
an unbiased importance sampling estimator
\[
\frac{Z\left(\theta\right)}{Z\left(\theta^{*}\right)}\approx\frac{\gamma\left(x\mid\theta\right)}{\gamma\left(x\mid\theta^{*}\right)},
\]
where $x\sim f\left(\cdot\mid\theta^{*}\right)$, which directly approximates
the ratio in equation \ref{eq:mcmc_Z} (an analogous approach is not
possible in IS). This estimator may also be improved using AIS, using
multiple importance points results in an inexact or ``noisy'' algorithm
\citep{Alquier2016}.
\item \citet{Grelaud2009} uses ABC, in which an unbiased estimate of the
approximate likelihood $\int_{x}f\left(y\mid\theta\right)\pi_{\epsilon}\left(y\mid x\right)dx$
is used, where $\pi_{\epsilon}$ is some kernel centred around the
data $y$. The ABC likelihood approaches the true likelihood as $\epsilon\rightarrow0$,
but in practice some small finite $\epsilon$ is used. In practice
the dimension of $y$ needs to be low for the likelihood estimate
to be accurate, thus usually summary statistics are used in place
of $y$ (introducing an additional approximation).
\end{enumerate}
All three approaches build likelihood estimators based on simulations
$x\sim f\left(\cdot\mid\theta^{*}\right)$. For most choices of doubly
intractable $f$ it is not possible to perform this simulation exactly.
\citet{Caimo2011} propose instead to use a long MCMC run, taking
the final point to be $x$, and \citet{Everitt2012} shows that the
bias this introduces goes to zero as the length of the MCMC run goes
to infinity. We refer to the exchange algorithm with this method of
generating $u$ as the ``approximate exchange algorithm''; the ``double
Metropolis-Hastings'' (DMH) sampler of \citet{Liang2010} is approximate
exchange, potentially with only a single MCMC step for generating
$x$.

The theme of this paper is in exploring the way in which simulations
from the likelihood may be used in the most computationally efficient
manner. We show how reusing simulations from the likelihood may lead
to efficient SMC samplers; then study the algorithms empirically.

\subsection{Outline}

\subsubsection{Motivation}

We are motivated by three separate aspects of performing inference
for doubly intractable distributions. This paper makes contributions
in each of these areas.
\begin{enumerate}
\item \textbf{Variance of likelihood estimators. }The pseudo-marginal approach
is an important method in the Bayesian computation toolbox, but can
perform poorly when the variance of the likelihood estimate is high.
Each of the three approaches in section \ref{subsec:Doubly-intractable-distributions}
can suffer from this problem. For example, in the SAV MCMC algorithm,
if $1/Z\left(\theta\right)$ is dramatically overestimated at a particular
$\theta$, the algorithm may get stuck for many iterations at this
$\theta$, even if it is in the tail of $\pi$. In such cases the
practical implication is that a very large sample of $\theta$ points
may be required in order to achieve low variance Monte Carlo estimates
with respect to $\pi\left(\theta\mid y\right)$.
\item \textbf{Population-based Monte Carlo for doubly intractable distributions.
}\citet{Caimo2011,Friel2013e} describe population-based MCMC methods
for Bayesian inference with doubly intractable distributions.\textbf{
}SMC samplers have proved a useful alternative to population-based
MCMC in a number of situations. \citet{Everitt2017} introduced a
random weight SMC sampler for doubly intractable distributions, but
it was restricted to the case of using a data point tempered sequence
of target distributions which may not always be suitable, and also
has the potential for bias to accumulate if the auxiliary variables
are not simulated exactly from $f\left(\cdot\mid\theta^{*}\right)$.
\item \textbf{Connecting ABC with auxiliary variable methods. }ABC has been
used as an alternative to exact auxiliary variable methods (e.g. \citet{Everitt2012}),
but its performance is not always competitive with these approaches
\citep{Friel2013e}. The link between ABC and auxiliary variable methods
has been remarked on before (both build a likelihood estimator based
on simulations $x\sim f\left(\cdot\mid\theta^{*}\right)$), but this
connection has not been explored more deeply.
\end{enumerate}

\subsubsection{Previous work and summary of contributions\label{subsec:Previous-work-and}}

In section \ref{sec:PMC-for-doubly} we build on work in \citet{Everitt2017}
to describe an alternative population-based Monte Carlo method for
inference in doubly intractable distributions. \citet{Everitt2017}
mentions the possibility of using marginal SMC, this being a very
similar algorithm to that in \citet{Koblents2015}, but does not discuss
the potential for this approach to overcome the limitations of the
other SMC algorithm described in that paper; i.e. using any sequence
of distributions is possible, and the accumulation of bias due to approximate
simulations of the auxiliary variables from $f\left(\cdot\mid\theta^{*}\right)$
may be avoided. In this paper we illustrate the benefits of these
properties, and go further by suggesting the design of an algorithm
that makes extensive use of previous iterations in order to lower
the variance of likelihood approximations.

In recent years there have been several approaches introduced to tackle
the issue of high variance likelihood estimates. One class of approaches
\citep{Dahlin2015a,Deligiannidis2016} uses coupling to introduce
a dependence between likelihood estimates for different values of
$\theta$, so that (for example) if an overestimation occurs at one
$\theta$ it also occurs for the next proposed value $\theta^{*}$,
thus balancing the numerator and the denominator in the acceptance
probability. Another class of approaches \citep{Wilkinson2014,Meeds2014a,Moores2015,Boland2016,Liang2015a,Sherlock2017}
pools the estimates for different $\theta$ and uses regression to
estimate the likelihood: this approach does not lead to unbiased estimates
of the likelihood and therefore is not exact, but the reduced variance
may lead to better performance for a fixed computational effort.

This paper introduces an approach that has elements in common with
\citet{Boland2016,Liang2015a} in that it adaptively makes use of
auxiliary variables that have been generated for previously visited
values of $\theta$. However, in contrast to these other approaches
it is straightforward to see that our method has the correct target:
\citet{Boland2016} construct a noisy MCMC algorithm that leads to
slightly biased estimates, and \citet{Liang2015a} relies on strong
mixing assumptions to prove that their adaptive MCMC algorithm converges
to the correct target.

In our case previously visited values of $\theta$ are values that
have been used in previous iterations of the SMC sampler. In many
cases the region of high posterior density in targets in the initial
iterations in the SMC covers the region of high posterior density
in later iterations. Thus we expect likelihood estimates at the $\theta$
used in previous iterations of the SMC to provide useful information
for subsequent iterations. It is important that, as achieved using
SMC, this population of existing $\theta$ should have a wider spread
than those in the true posterior; if this is not the case, then it
is likely that there will be very few existing $\theta$ in the tails
of the posterior (such as may be encountered in \citet{Sherlock2017}),
leading to high variance likelihood estimators in these regions. When
used within an MCMC method we may expect that this leads the chain
to be prone to get stuck in the tails of the posterior. Related approaches
use different methods for constructing a useful population of existing
$\theta$, which involve running a different algorithm as a preliminary
step: in \citet{Boland2016} this population is constructed using
a Laplace approximation to the posterior; in \citet{Liang2015a} this
is constructed using DMH (with ABC being suggested as an alternative
that provides wider support than the true posterior). Outside of work
on doubly intractable distributions, \citet{South2016} describes
an approach for reusing points from earlier SMC iterations.

In section \ref{subsec:Auxiliary-variable-marginal} we describe our
SMC algorithm, and in section \ref{subsec:Choosing-the-path} we present
a novel approach to automatically deciding which previous values of
$\theta$ to use in the likelihood estimate. Then in section \ref{sec:Empirical-results}
we present empirical results, in which we compare the new method with
a range of exisiting techniques, applied to the Ising model, before
a concluding discussion in section \ref{sec:Discussion}. Two of the
methods we compare empirically are the auxiliary variable approaches
of \citet{Moller2006} and ABC. In appendix \ref{sec:Connecting-ABC-with}
we show that these method can be seen to be related through introducing
a novel derivation of the MAV method arrived at by aiming to improve
the standard ABC approach where the full data is used. This connection
between ABC and auxiliary variable methods serves to highlight why
we would expect certain types of ABC approach to exhibit worse performance
than auxiliary variable methods.

\section{Marginal SMC for doubly intractable distributions\label{sec:PMC-for-doubly}}

\subsection{SMC samplers with estimated likelihoods\label{subsec:SMC-samplers-with}}

\citet{Sisson2007f,Beaumont2009,Fearnhead2010a,DelMoral2012,Chopin2013,Moores2015,Koblents2015,Everitt2017}
all describe exact-approximate methods for using estimated likelihoods
within SMC. In each case, an auxiliary variable construction may be
used to provide the correctness of the algorithm: in most cases the
SMC algorithm can be seen to be an instance of \citet{DelMoral2006c}
in an extended space. However, when using an estimated likelihood,
it is easy to see that some very standard configurations of SMC samplers
are no longer available. We now give an example of this.

The most fundamental choice is of the sequence of distributions $\pi_{t}$.
In some cases there are clear choices for this: for example in ABC,
where the sequence of distributions uses the ABC approximation with
a decreasing sequence of $\epsilon_{t}$, so that there is a move
from approximate targets that are easily computed and close to the
prior towards more accurate targets that are close to the posterior
but are more difficult to estimate. One general way of helping the
sampler to easily locate regions of high posterior mass is to use
some sort of geometric annealing: a natural choice is to use $\pi_{t}\left(\theta\right)=p\left(\theta\right)\hat{f}^{\nu_{t}}\left(y|\theta\right)$,
where $\hat{f}^{\nu_{t}}$ is the estimated likelihood raised to a
power and $\nu_{t}$ moves from 0 to 1 as $t$ increases. This choice
has the additional benefit in this situation of allowing values of
$\hat{f}$ calculated at each iteration of the SMC sampler to be used
in future iterations of the sampler, via the ideas of regression and
pre-computation mentioned in section \ref{subsec:Previous-work-and}.

However, we note that even if $\hat{f}$ is an unbiased estimate of
$f$, $\hat{f}^{\nu_{t}}$ is not an unbiased estimate of $f^{\nu_{t}}$,
thus it is not immediately obvious how to construct a random weight
SMC sampler, using this sequence of distributions, that results in
the correct target distribution. For example, if we consider the case
where the kernel $K_{t}$ is chosen to be an MCMC kernel targeting
$\pi_{t}$, if the likelihood is directly available we obtain the
(unnormalised) weight of particle $p$ at target $t$ as
\[
\tilde{w}_{t}^{(p)}=\frac{p\left(\theta_{t-1}^{(p)}\right)f^{\nu_{t}}\left(y|\theta_{t-1}^{(p)}\right)}{p\left(\theta_{t-1}^{(p)}\right)f^{\nu_{t-1}}\left(y|\theta_{t-1}^{(p)}\right)}w_{t-1}^{(p)}=f^{\left(\nu_{t}-\nu_{t-1}\right)}\left(y|\theta_{t-1}^{(p)}\right)w_{t-1}^{(p)},
\]
where $w_{t-1}^{(p)}$ is the normalised weight of particle $p$ at
iteration $t-1$. Using the weight 
\[
\tilde{w}_{t}^{(p)}=\hat{f}^{\left(\nu_{t}-\nu_{t-1}\right)}\left(y|\theta_{t-1}^{(p)}\right)w_{t-1}^{(p)}
\]
as a direct substitute for this yields a biased estimate of the weight.
One might hope that this bias may not be important in practice, since
this weight update implicitly specifies a sequence of target distributions
(that is different from $f^{\nu_{t}}$), however it is not then in
general possible to give an MCMC kernel that has the implicit target
as its invariant distribution. Therefore this algorithm may be considered
as an example of a noisy SMC algorithm (which we note, if a similar
substitution is made in the MCMC kernel acceptance probability, uses
a noisy MCMC update) of the type studied in \citet{Everitt2017}.
This paper notes that the use of biased weights leads to bias accumulating,
with the approximate target potentially not being close to the desired
target. One way if avoiding this problem would be to use to debias
the weight estimates (e.g. \citet{Girolami2013}) however we do not
pursue this idea here due to the high computational expense of these
approaches.

\subsection{Marginal SMC with estimated likelihoods\label{subsec:Marginal-SMC-with}}

There is a simple way to use the annealed sequence of target distributions
that results in the true target. That is to use the marginal SMC algorithm,
in which the optimal SMC sampler backward kernel at target $t$ is
approximated using the empirical distribution of the particles at
iteration $t-1$. This algorithm is very similar to population Monte
Carlo \citep{cappe2004population} (the difference is that in the
SMC case we have the target distribution changing as the algorithm
iterates). The pseudo-marginal PMC algorithm proposed in \citet{Koblents2015}
is thus very similar to the marginal SMC approach here, and the procedure
described in that paper for regularising the IS weights at each iteration
is not dissimilar to changing the sequence of targets as the algorithm
progresses.

The weight update at target $t$ in this case will be
\begin{equation}
\tilde{w}_{t}^{(p)}=\frac{p\left(\theta_{t}^{(p)}\right)\hat{f}^{\nu_{t}}\left(y|\theta_{t}^{(p)}\right)}{\sum_{r=1}^{P}w_{t-1}^{(r)}K_{t}\left(\theta_{t}^{(p)}\mid\theta_{t-1}^{(r)}\right)}\label{eq:pmc_weight}
\end{equation}
For $\gamma_{t}<1$ this weight is a biased estimate of the weight
that uses the true $f$, thus this update does not result in a target
distribution at iteration $t$ of $p(\theta)f^{\nu_{t}}\left(y|\theta\right)$.
However, when $\nu_{t}=1$ (say for $t=T$), the weight is an unbiased
estimate, thus by the auxiliary variable argument in section \ref{subsec:The-pseudo-marginal-approach},
with proposal
\[
q\left(\theta_{t}^{(p)}\right)=\sum_{r=1}^{P}w_{t-1}^{(r)}K_{t}\left(\theta_{t}^{(p)}\mid\theta_{t-1}^{(r)}\right)
\]
this scheme yields the exact target, with an unbiased estimate of
the marginal likelihood being given by
\begin{equation}
\hat{p}\left(y\right)=\frac{1}{P}\sum_{p=1}^{P}\tilde{w}_{T}^{(p)}.\label{eq:ml_est_pmc}
\end{equation}
The marginal SMC algorithm has the important property in this context
that at every SMC iteration, it integrates over the previous target,
rather than sampling from the path space of targets as a standard
SMC sampler does. This is the reason that bias does not accumulate
this algorithm: it is essentially an IS algorithm with a well-tuned
proposal. A common choice for $K_{t}\left(\cdot\mid\theta_{t-1}\right)$
is a Gaussian with mean $\theta_{t-1}$; the variance may be chosen
adaptively, e.g. \citet{Beaumont2009} suggests to choose the variance
to be twice the weighted sample variance of the current particles
$\left\{ \theta_{t-1}^{(p)}\right\} _{p=1}^{P}$.

This style of algorithm (without the geometric annealing and with
a different sequence of targets), in the guise of population Monte
Carlo (PMC) has been popular for exploring ABC posteriors \citep{Beaumont2009}.
There are a few disadvantages of using this approach rather than standard
SMC. Firstly, each iteration has cost $\mathcal{O}\left(P^{2}\right)$
compared to $\mathcal{O}\left(P\right)$ for standard SMC (although
\citet{Klaas2005} notes that this can be reduced to $\mathcal{O}\left(P\log\left(P\right)\right)$
with little loss in accuracy). Secondly, since the method is an importance
sampler (without any MCMC moves as in standard SMC) it is unlikely
to be efficient in high ($>50$, say) dimensional parameter spaces.

\subsection{Auxiliary variable marginal SMC for doubly intractable distributions\label{subsec:Auxiliary-variable-marginal}}

\subsubsection{Auxiliary variable marginal SMC\label{subsec:Standard-PMC}}

We now return to the SAV estimator of the likelihood of a doubly intractable
distribution from section \ref{subsec:Doubly-intractable-distributions}
that uses
\begin{equation}
\hat{f}\left(y\mid\theta\right)=\frac{\gamma\left(y\mid\theta\right)}{Z\left(\hat{\theta}\right)}\widehat{\frac{Z\left(\hat{\theta}\right)}{Z\left(\theta\right)}}=\frac{\gamma\left(y\mid\theta\right)}{Z\left(\hat{\theta}\right)}\frac{\gamma\left(x\mid\hat{\theta}\right)}{\gamma\left(x\mid\theta\right)},\label{eq:sav_est}
\end{equation}
with $x\sim f\left(\cdot\mid\theta\right)$. We propose to use a marginal
SMC algorithm with a sequence of $T$ targets using, at the $t$th
target the weight update in equation \ref{eq:pmc_weight} where the
estimated likelihood is that in equation \ref{eq:sav_est}. The variance
of the weights is dependent on the choice of $\hat{\theta}$; to minimise
the variance a reasonable choice for $\hat{\theta}$ is a value that
is close to many of the $\theta^{(p)}$, such as a maximum likelihood
estimate. When using a sequence of targets, there is no reason to
expect a single value of $\hat{\theta}$ to be appropriate for every
target, therefore we propose to alter this value at each marginal
SMC iteration, using $\hat{\theta}_{t}$ at the $t$th iteration.
The use of marginal SMC allows us to base the choice of $\hat{\theta}_{t}$
on the set of particles simulated at target $t-1$: we choose
\[
\hat{\theta}_{t}=\sum_{p=1}^{P}w_{t-1}^{(p)}\theta_{t-1}^{(p)}.
\]
We note that for parameter estimation, it is sufficient to use the
weights without the term $1/Z\left(\hat{\theta}_{t}\right)$, however
for obtaining an unbiased estimate of the evidence we would require
to multiply the expression in equation \ref{eq:ml_est_pmc} by an
unbiased estimate $\widehat{1/Z\left(\hat{\theta}_{t}\right)}$.

The complete algorithm is given in algorithm \ref{alg:Auxiliary-variable-marginal}.
In the remainder of this section we describe improvements to this
basic algorithm that make further use of the set of $\theta$ from
previous iterations.

\begin{algorithm}
\begin{algorithmic}
\For {$p= 1:P$}
    \State $\theta^{(p)}_0 \sim p\left( \cdot \right)$
    \State $x^{(p)}_{0} \sim f\left( \cdot \mid \theta^{(p)}_0 \right)$
\EndFor
\State $t=0$.
\For {$t=0:T-1$}
    \State $\hat{\theta}_{t+1}=\sum_{p=1}^{P}w_{t}^{(p)}\theta_{t}^{(p)}$
    \For {$p= 1:P$}
        \State $\theta^{(p)}_{t+1} \sim K_{t+1}\left( \cdot \mid \theta^{(p)}_{t} \right)$
        \State $x^{(p)}_{t+1} \sim f\left( \cdot \mid \theta^{(p)}_{t+1} \right)$
    \EndFor
    \For {$p= 1:P$}
        \State $\tilde{w}_{t+1}^{(i)}=\frac{p\left(\theta_{t+1}^{(p)}\right)\gamma^{\nu_{t+1}}\left(y|\theta_{t+1}^{(p)}\right) }{ \sum_{r=1}^{P}w_{t}^{(r)}K_{t+1}\left(\theta_{t+1}^{(p)}\mid\theta_{t}^{(r)}\right)} \left( \frac{\gamma \left( x^{(p)}_{t+1} \mid\hat{\theta}_{t+1}\right)}{\gamma \left( x^{(p)}_{t+1} \mid \theta_{t+1}^{(p)} \right)} \right)^{\nu_{t+1}}$
    \EndFor
    \State Normalise $\left\{ \tilde{w}_{t+1} \right\}_{i=1}^N$ to give normalised weights $\left\{ w_{t+1} \right\}_{i=1}^N$.
    \State Resample.
\EndFor
\end{algorithmic}

\caption{Auxiliary variable marginal SMC with $M=1$.\label{alg:Auxiliary-variable-marginal}}

\end{algorithm}

\subsubsection{Path marginal SMC\label{subsec:Path-marginal-SMC}}

\citet{Boland2016} introduce a pre-computation scheme for improving
the estimation of ratios of the partition function the regions of
the MAP estimate of $\theta$. The procedure uses the following steps.
\begin{enumerate}
\item Locate the MAP estimate of $\theta$ using a stochastic approximation
algorithm.
\item Estimate the Hessian at the MAP estimate, and use this to construct
a grid of points designed to cover the region where most of the posterior
mass is found.
\item At every point on the grid, simulate from the likelihood: at the $j$th
point in the grid, simulate $x^{(j)}\sim f\left(\cdot\mid\theta^{(j)}\right)$.
\item Use these simulations in estimates of ratios of the partition function.
\end{enumerate}
\citet{Boland2016} make use of this grid of points to create a variation
on an MCMC algorithm used for simulating from doubly intractable distributions;
the exchange algorithm. In the exchange algorithm, the true (but intractable)
acceptance probability
\[
\alpha\left(\theta,\theta^{*}\right)=\frac{p\left(\theta^{*}\right)\gamma\left(y\mid\theta^{*}\right)Z\left(\theta\right)}{p\left(\theta\right)\gamma\left(y\mid\theta^{*}\right)Z\left(\theta^{*}\right)}
\]
is replaced with the acceptance probability
\[
\alpha\left(\theta,\theta^{*}\right)=\frac{p\left(\theta^{*}\right)\gamma\left(y\mid\theta^{*}\right)\gamma\left(x\mid\theta\right)}{p\left(\theta\right)\gamma\left(y\mid\theta^{*}\right)\gamma\left(x\mid\theta^{*}\right)},
\]
where $x\sim f\left(\cdot\mid\theta^{*}\right)$. One may view this
algorithm as using $\gamma\left(x\mid\theta\right)/\gamma\left(x\mid\theta^{*}\right)$
as an estimator of $Z\left(\theta\right)/Z\left(\theta^{*}\right)$.
Despite the use of an estimate of the true acceptance probability,
this method results in targeting the correct posterior distribution,
although the same is not true of most other methods that use an estimate
of $Z\left(\theta\right)/Z\left(\theta^{*}\right)$ (even when this
estimate is unbiased). The variation used by \citet{Boland2016} is to
use the estimate
\begin{eqnarray*}
\frac{Z\left(\theta\right)}{Z\left(\theta^{*}\right)} & = & \frac{Z\left(\theta\right)}{Z\left(\theta_{a}\right)}\times\frac{Z\left(\theta_{a}\right)}{Z\left(\theta_{b}\right)}\times\frac{Z\left(\theta_{b}\right)}{Z\left(\theta^{*}\right)}\\
 & \approx & \widehat{\frac{Z\left(\theta\right)}{Z\left(\theta_{a}\right)}}\times\widehat{\frac{Z\left(\theta_{a}\right)}{Z\left(\theta_{b}\right)}}\times\left(\widehat{\frac{Z\left(\theta^{*}\right)}{Z\left(\theta_{b}\right)}}\right)^{-1},
\end{eqnarray*}
where $\theta_{a}$ is the nearest point in the grid to $\theta$
and $\theta_{b}$ is the nearest to $\theta^{*}$ and each of the
estimates is provided by importance sampling in the manner used throughout
this paper, noting that since $\theta_{a}$ and $\theta_{b}$ are
in the precomputed grid, the necessary simulation from $f\left(\cdot\mid\theta_{a}\right)$
and $f\left(\cdot\mid\theta_{b}\right)$ has been performed prior
to running the MCMC algorithm. Although this does not result in a
unbiased estimate of $Z\left(\theta\right)/Z\left(\theta^{*}\right)$
and the algorithm does not target the correct posterior distribution,
due to the precomputation the method is extremely fast, and \citet{Boland2016} show that this method is an instance of a \emph{noisy MCMC}
algorithm and are able to provide some theoretical guarantees as to
its convergence. We might expect such this method to perform well
when the precomputed grid of points provides reasonable coverage of
the posterior. A similar scheme is used in the Hamiltonian Monte Carlo
approach in \citet{Stoehr2017a}.

In this paper we consider the use of a similar approach within the
marginal SMC method described in section \ref{subsec:Standard-PMC}.
Rather than using a precomputed grid of points the idea in what we
call \emph{path marginal SMC} is to make use, in a similar way to
that described above, of simulations performed for previous populations
of particles. For a tempered sequence of targets, we expect that particles
from previous iterations cover the the support of the target at the
current iteration.

In algorithm \ref{alg:Auxiliary-variable-marginal}, at the $t$th
iteration, we first simulate a new population of points $\theta_{t}:=\left\{ \theta_{t}^{(p)}\right\} _{p=1}^{P}$,
then simulate a corresponding population of auxiliary variables $x_{t}:=\left\{ x_{t}^{(p)}\right\} _{p=1}^{P}$.
Path marginal SMC uses follows exactly these steps, then uses an alternative
method for calculating the weights, specifically, using a different
method to estimate the term $Z\left(\hat{\theta}\right)/Z\left(\theta_{t}^{(p)}\right)$.
We propose to use the product
\begin{eqnarray}
\frac{Z\left(\hat{\theta}_{t}\right)}{Z\left(\theta_{t}^{(p)}\right)} & = & \frac{Z\left(\theta_{\Pi_{1}}\right)}{Z\left(\theta_{\Pi_{0}}\right)}\times\ldots\times\frac{Z\left(\theta_{\Pi_{l}}\right)}{Z\left(\theta_{\Pi_{l-1}}\right)},\label{eq:path_Z}
\end{eqnarray}
where $\theta_{\Pi_{0}},...,\theta_{\Pi_{l}}$ are $l+1$ points on
a path $\Pi$ chosen from the points $\theta_{1:t}$ visited over
the entire history of the marginal SMC thus far, with $\Pi_{0}=\theta_{t}^{(p)}$
and $\Pi_{l}=\hat{\theta}_{t}$. We may estimate each term in this
product by importance sampling, making use of the previously simulated
auxiliary variables corresponding to each of the $\theta_{\pi_{i}}$
in the path. to construct lower variance estimators (similarly to
\citet{Friel2013e}, and also \citet{Boland2016,Stoehr2017a}). If
the path is chosen appropriately, this will provide a lower variance
estimate of $Z\left(\hat{\theta}_{t}\right)/Z\left(\theta_{t}^{(p)}\right)$
than that used in standard marginal SMC, in the same manner as using
an AIS estimate. The estimate of $Z\left(\hat{\theta}_{t}\right)/Z\left(\theta_{t}^{(p)}\right)$
is unbiased, thus it is easy to see that the importance points end
up being weighted according to the correct target distribution (the
estimates being correlated does not affect this result). We note that
using a similar scheme in MCMC (where we choose a path from the past
history of the chain, as in the method in \citet{Sherlock2017}) would
not result in the correct target distribution, since making use of
auxiliary variables from before the previous iteration would break
the Markov assumption, however in marginal SMC we do not face such
a restriction.

To be specific, we use the estimator
\begin{equation}
\frac{Z\left(\theta_{\Pi_{i+1}}\right)}{Z\left(\theta_{\Pi_{i}}\right)}\approx\frac{1}{M}\sum_{m=1}^{M}\frac{\gamma\left(x_{\Pi_{i}}^{(m)}\mid\theta_{\Pi_{i+1}}\right)}{\gamma\left(x_{\Pi_{i}}^{(m)}\mid\theta_{\Pi_{i}}\right)},\label{eq:path_est_each}
\end{equation}
where $x_{\Pi_{i}}^{(m)}\sim f\left(\cdot\mid\theta_{\Pi_{i}}\right)$
from some previous iteration of the marginal SMC. For simplicity we
describe the case where $M=1$, so that the the estimator is
\begin{equation}
\hat{R}:=\widehat{\frac{Z\left(\hat{\theta}_{t}\right)}{Z\left(\theta_{t}^{(p)}\right)}}=\prod_{i=0}^{l-1}\frac{\gamma\left(x_{\Pi_{i}}\mid\theta_{\Pi_{i+1}}\right)}{\gamma\left(x_{\Pi_{i}}\mid\theta_{\Pi_{i}}\right)}.\label{eq:path_est}
\end{equation}

For many distributions, in order to evaluate $\gamma\left(\cdot\mid\theta\right)$
we need only a low dimensional statistic $S\left(x\right)$ of $x$.
Therefore after the $x$ values are simulated, we need only store
this statistic for use in future iterations. In order for the scheme
to be most effective we require that the convex hull of the points
simulated from target $t$ contains the convex hull of the points
from target $t+1$.

Our approach also yields an estimate of the marginal likelihood, when
combined with an estimate for $1/Z\left(\hat{\theta}_{t}\right)$.
The method is summarised in algorithm \ref{alg:Path-marginal-SMC.}.

\begin{algorithm}
\begin{algorithmic}
\For {$p= 1:P$}
    \State $\theta^{(p)}_0 \sim p\left( \cdot \right)$
    \State $x^{(p)}_{0} \sim f\left( \cdot \mid \theta^{(p)}_0 \right)$
\EndFor
\State $t=0$.
\For {$t=0:T-1$}
    \State $\hat{\theta}_{t+1}=\sum_{p=1}^{P}w_{t}^{(p)}\theta_{t}^{(p)}$
    \For {$p= 1:P$}
        \State $\theta^{(p)}_{t+1} \sim K_{t+1}\left( \cdot \mid \theta^{(p)}_{t} \right)$
        \State $x^{(p)}_{t+1} \sim f\left( \cdot \mid \theta^{(p)}_{t+1} \right)$
    \EndFor
    \For {$p= 1:P$}
        \State Find a path $\Pi$ between $\theta_{t}^{(p)}$ and $\hat{\theta}_{t+1}$ using the method in section \ref{subsec:Choosing-the-path}.
        \State $\tilde{w}_{t+1}^{(i)}=\frac{p\left(\theta_{t+1}^{(p)}\right)\gamma^{\nu_{t+1}}\left(y|\theta_{t+1}^{(p)}\right) }{ \sum_{r=1}^{P}w_{t}^{(r)}K_{t+1}\left(\theta_{t+1}^{(p)}\mid\theta_{t}^{(r)}\right)} \left( \prod_{i=0}^{l-1}\frac{\gamma\left(x_{\Pi_{i}}\mid\theta_{\Pi_{i+1}}\right)}{\gamma\left(x_{\Pi_{i}}\mid\theta_{\Pi_{i}}\right)} \right)^{\nu_{t+1}}$
    \EndFor
    \State Normalise $\left\{ \tilde{w}_{t+1} \right\}_{i=1}^N$ to give normalised weights $\left\{ w_{t+1} \right\}_{i=1}^N$.
    \State Resample.
\EndFor
\end{algorithmic}

\caption{Path marginal SMC.\label{alg:Path-marginal-SMC.}}
\end{algorithm}

\subsection{Choosing the path\label{subsec:Choosing-the-path}}

The variance of the estimator based on equation \ref{eq:path_Z} depends
on the choice of path $\Pi$. The estimator has the same character
as that in AIS, and in section \ref{subsec:Variance-of-path} we extend
the arguments of \citet{Neal2001} to (under some assumptions) give
its variance for any path. We see that this expression provides a
simple criterion with which we may evaluate the likely variance given
by different paths. Then in section \ref{subsec:Searching-over-paths}
we describe how to efficiently search over the space of possible paths
in order to locate a path that provides a low variance estimator.

\subsubsection{Variance of path estimator\label{subsec:Variance-of-path}}

To study the variance of the path estimator (which a product of IS
estimators), we study the variance of the $\log$ of equation \ref{eq:path_est}
as in \citet{Neal2001}. We have
\[
\log\left(\hat{R}\right)=\sum_{i=0}^{l-1}\log\gamma\left(x_{i}\mid\theta_{\Pi_{i+1}}\right)-\log\gamma\left(x_{i}\mid\theta_{\Pi_{i}}\right).
\]
Now suppose that $f$ is in the exponential family with natural parameter
$\theta$, so that
\[
f\left(x\mid\theta\right)=\frac{\gamma\left(x\mid\theta\right)}{Z\left(\theta\right)}=\frac{h\left(x\right)\exp\left(\theta^{T}S\left(x\right)\right)}{Z\left(\theta\right)}
\]
In this case, for any $\theta_{A}$, $\theta_{B}$, 
\[
\log\gamma\left(x\mid\theta_{A}\right)-\log\gamma\left(x\mid\theta_{B}\right)=\left(\theta_{A}-\theta_{B}\right)^{T}S\left(x\right).
\]
Let $\mathbb{V}_{\Pi_{i}}$ be the variance under $f\left(\cdot\mid\theta_{\Pi_{i+1}}\right)$
and $\mathbb{V}_{\Pi_{1}^{l-1}}$ be the variance under the path of
distributions given by $\Pi$. Then
\begin{eqnarray}
\mathbb{V}_{\Pi_{1}^{l-1}}\left[\log\left(\hat{R}\right)\right] & = & \sum_{i=0}^{l-1}\mathbb{V}_{\Pi_{i}}\left[\left(\theta_{\Pi_{i+1}}-\theta_{\Pi_{i}}\right)^{T}S\left(x_{i}\right)\right]\nonumber \\
 & = & \sum_{i=0}^{l-1}\left(\theta_{\Pi_{i+1}}-\theta_{\Pi_{i}}\right)^{T}\mathbb{V}_{\Pi_{i}}\left[S\left(x_{i}\right)\right]\left(\theta_{\Pi_{i+1}}-\theta_{\Pi_{i}}\right)\label{eq:complex_score}
\end{eqnarray}
An estimate of this quantity may be used as a means for comparing
the variance of estimators of the form in equation \ref{eq:path_est}.
To highlight the connection with AIS we make two simplifying assumptions:
that the covariance $V_{i}=\mathbb{V}_{\Pi_{i}}\left[S\left(x_{i}\right)\right]$
is constant over $i$ and that $V$ is diagonal, so that the covariance
between different dimensions of $S\left(x_{i}\right)$ is zero, then
\begin{eqnarray}
\mathbb{V}_{\Pi_{1}^{l-1}}\left[\log\left(\hat{R}\right)\right] & = & \left[\sum_{i=0}^{l-1}\left(\theta_{\Pi_{i+1}}-\theta_{\Pi_{i}}\right)^{2}\right]^{T}\text{diag}\left[V\right].\label{eq:score}
\end{eqnarray}
Under these assumptions, the sum of weighted (by $V$) squared distances
between the successive parameters in the path determines the efficiency
of the estimator. From this we see that the estimator used in AIS,
in which the path forms a straight line between the parameters at
the start and end of the path, is optimal using this criterion, and
that its variance is inversely proportional to $l$. In our SMC scheme,
such a path is not available to us, but we see that we may improve
on the IS estimator (where $l=1$) by using a path that is sufficiently
close to the straight line between the end points. Having the support
of each target distribution being contained in the support of the
previous target distribution makes it more likely that such a path
may exist, although when $\theta$ is high dimensional the probability
of finding a path close to this line is likely to be low (also noting
that similar issues to those encountered in nearest neighbour methods
\citep{Aggarwal2001} in high dimensions also apply here). This limits
the use of our approach to problems of low to moderate dimension,
but since the path estimator is embedded in marginal SMC, we have
already restricted our attention to these cases. However this is not
overly restrictive, since in the most common instances of doubly intractable
distributions, the parameter space is less than 10-dimensional.

\subsubsection{Searching over paths\label{subsec:Searching-over-paths}}

To use the path estimator, at the $t$th iteration of the SMC, for
every particle $\theta_{t}^{(p)}$, we need to choose an appropriate
path between $\theta_{t}^{(p)}$ and $\hat{\theta}_{t}$, We use the
sum of weighted squared distances from equation \ref{eq:complex_score}
to score each path, but we cannot in practice search exhaustively
through all possible paths since the number of these grows factorially
with the number of points available to choose from. In this section
we describe a heuristic approach, in which we first restrict our attention
to points close to $\theta_{t}^{(p)}$ and $\hat{\theta}_{t}$, then
greedily search for the optimal path amongst these points. Our greedy
algorithm is motivated by aiming to restrict attention to paths close
to the straight line between $\theta_{t}^{(p)}$ and $\hat{\theta}_{t}$,

To begin we require an estimate of $V_{i}$. For this we assume that
$V_{i}$ is the same for all $i$, and take $\hat{V}$ to be the empirical
variance of $\left\{ S\left(x_{t}^{(p)}\right)\right\} _{p=1}^{P}$,
where each $x_{t}^{(p)}\sim f\left(\cdot\mid\theta_{t}^{(p)}\right)$
has already been simulated. This estimate may be quite inaccurate
in the initial stages of the algorithm since it relies on the assumption
that the mean of $S\left(x\right)$ is the same over all $\left\{ \theta_{t}^{(p)}\right\} _{p=1}^{P}$.
However, as the algorithm progresses, the points $\left\{ \theta_{t}^{(p)}\right\} _{p=1}^{P}$
become closer together, making the assumptions of a constant mean
and variance of $S\left(x\right)$ more appropriate over this set
of parameters.

\begin{figure}
\begin{centering}
\includegraphics[scale=0.8]{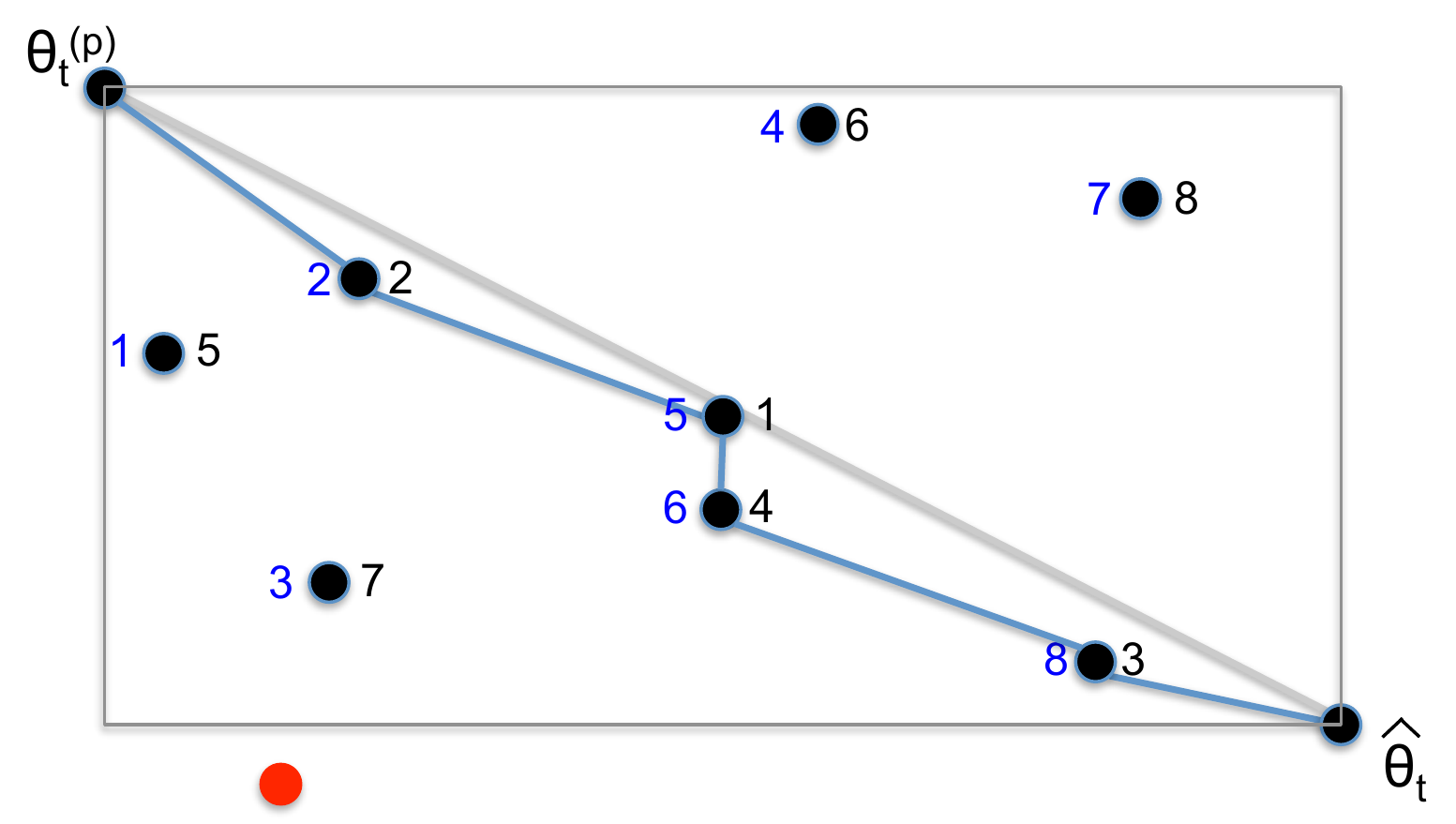}
\par\end{centering}
\caption{A sketch of the algorithm used to find the path used in the estimator.
Points inside of the minimum bounding box between $\theta_{t}^{(p)}$
and $\hat{\theta}_{t}$ are shown in black, and a point outside is
shown in red. The points are numbered both in the order of distance
to the straight line between $\theta_{t}^{(p)}$ and $\hat{\theta}_{t}$
(in black) from step 2 and in order of the list $\mathcal{L}$ from
step 3 (in blue). The resultant path is shown in dark blue.\label{fig:A-sketch-of}}

\end{figure}

For target $t$ and particle $p$, our approach takes the following
steps (summarised in figure \ref{fig:A-sketch-of}).
\begin{enumerate}
\item \textbf{Range searching.} Let $B$ be the minimum bounding box of
$\theta_{t}^{(p)}$ and $\hat{\theta}_{t}$. Our first step is to
locate all points in $\left\{ \theta_{s}^{(p)}\right\} _{s=1:t,p=1:P}$
that fall in $B$: it is sufficient to restrict our attention to these
points since points outside of this box are guaranteed to increase
the sum of squared differences. This may be accomplished efficiently
using a KD-tree \citep{Bentley1975} (which is the same for every
particle), as is also used in \citet{Sherlock2017} for a different
purpose. Finding points in a box is called a ``range search'', and
this has a cost of $O\left(d+\log\left(tP\right)\right)$. To avoid
building the tree at each SMC iteration (which has cost $O\left(tP\log\left(tP\right)\right)$),
it is possible to update the tree from iteration $t-1$ with the $P$
new points from iteration $t$ with cost approximately $O\left(\log\left(tP\right)\right)$)
per point. Eventually a tree constructed using such a scheme will
need ``rebalancing'' in order to keep the cost of the range search
low. If the cost of constructing the tree becomes too large, we may
restrict our attention to locating points from the past $L$ target
distributions rather than the whole history of the SMC.
\item \textbf{Sorting.} Suppose that $Q$ points fall inside $B$ (not including
$\theta_{t}^{(p)}$ and $\hat{\theta}_{t}$). First we find the Euclidean
distance between each of the points in $B$ and: $\theta_{t}^{(p)}$;
$\hat{\theta}_{t}$; and the straight line between $\theta_{t}^{(p)}$
and $\hat{\theta}_{t}$. We then sort the points in $B$ in order
of the distance to the straight line between $\theta_{t}^{(p)}$ and
$\hat{\theta}_{t}$.. The sorted list of points is denoted by $\left(\vartheta_{q}\right)_{q=1}^{Q}$,
with $\vartheta_{0}=\theta_{t}^{(p)}$. This sort operation has complexity
$O\left(Q\log\left(Q\right)\right)$, which is dominated by the cost
of the range searching step, especially since $Q$ is likely to be
small.
\item \textbf{Restricting the space of paths.} We create two sorted lists
of the points in $B$: $\mathcal{L}_{1}$ in order of increasing distance
to $\theta_{t}^{(p)}$; and $\mathcal{L}_{2}$ in order of decreasing
distance to $\hat{\theta}_{t}$. We then assign a score to each point
in $B$ by adding the index of the point in $\mathcal{L}_{1}$ to
the index of the point in $\mathcal{L}_{2}$. The points in $B$ are
then sorted in increasing order of this score (in list $\mathcal{L}$),
and we restrict our attention to paths that pass through the points
in this order. This step has complexity $O\left(Q\log\left(Q\right)\right)$.
\item \textbf{Growing the path. }We may use the following greedy algorithm,
which has cost $O\left(Q\right)$, to search through the space of
possible paths. We initialise the algorithm with the path $\Pi$ that
moves directly from $\theta_{t}^{(p)}$ to $\hat{\theta}_{t}$ and
assign it the score $D=\left(\hat{\theta}_{t}-\theta_{t}^{(p)}\right)^{T}\hat{V}\left(\hat{\theta}_{t}-\theta_{t}^{(p)}\right)$.
For $q=1:Q$ we perform the following steps:
\begin{enumerate}
\item Let $\vartheta_{c}^{*}=\vartheta_{q}$. Add $\vartheta_{c}^{*}$ to
the existing path in the position dictated by step 3, giving the proposed
path $\Pi^{*}$ and evaluate the score $D^{*}$ for this path using
equation \ref{eq:complex_score} with $\hat{V}$ in place of $\mathbb{V}_{\Pi_{i}}\left[S\left(x_{i}\right)\right]$. 
\item If $D^{*}<D$, then let $\Pi=\Pi{}^{*}$ and $D=D^{*}$.
\end{enumerate}
\end{enumerate}

We will see empirically that, although this approach will not necessarily
find the optimal path (except in one dimension, where it yields the
path containing all of the points along the line between $\theta_{t}^{(p)}$
and $\hat{\theta}_{t}$), it leads to low variance path estimates.

There are several possible extensions, most notably that if we do
not assume that $\mathbb{V}_{\Pi_{i}}\left[S\left(x_{i}\right)\right]$
is constant over $i$, for some models this quantity may be estimated
for each $i$ by means of a bootstrap (for example, the block bootstrap
for the Ising model) that recycles existing simulations. We note also
that since estimates based on different paths are unbiased, we may
find estimates based on several different paths and take their mean
to combine them and achieve a lower variance.

\section{Empirical results\label{sec:Empirical-results}}

\subsection{Ising model}

In this section we compare the result of running several different
algorithms on a previously studied $10\times10$ node two-dimensional
Ising model. An Ising model is a pairwise Markov random field model
on binary variables, each taking values in $\left\{ -1,1\right\} $.
The data is taken from \citet{Friel2013e} and, as in that paper we
consider two different neighbourhood structures. In both cases, the
variables are arranged in a grid: in the first order model, the neighbours
of each node are the nodes horizontally and vertically adjacent to
it, with $\mathbf{N}_{1}$ denoting the set of pairs of such neighbours;
in the second order model a second set of neighbours are additionally
used, with $\mathbf{N}_{2}$ denoting the set of pairs of these neighbours.
The distributions of the first and second order models are respectively
\[
l_{1}\left(y\mid\theta_{1}\right)\propto\exp\left(\theta_{1}S_{1}\left(y\right)\right),
\]
and
\[
l_{2}\left(y\mid\theta_{1},\theta_{2}\right)\propto\exp\left(\theta_{1}S_{1}\left(y\right)+\theta_{2}S_{2}\left(y\right)\right),
\]
where $\theta_{1},\theta_{2}\in\mathbb{R}$, $y_{i}$ denotes the
$i$th random variable in $y$ and $S_{1}\left(y\right)=\sum_{\left(i,j\right)\in\mathbf{N}_{1}}y_{i}y_{j}$,
$S_{2}\left(y\right)=\sum_{\left(i,j\right)\in\mathbf{N}_{2}}y_{i}y_{j}$.

We ran the following algorithms 40 times on both models: the exchange
algorithm; SAV-MCMC; ABC-MCMC; auxiliary variable marginal SMC (SAV-mSMC)
(as in section \ref{subsec:Standard-PMC}), and path marginal SMC
(path-mSMC) (section \ref{subsec:Path-marginal-SMC}) and recorded
the estimated posterior expectation and (co)variance in each case. All
algorithms were run using $100$ sweeps of a single-site update Gibbs
sampler to simulate from the likelihood, taking the final point as
the simulated $x$ variable. The same computational budget, measured
by the number of simulations from the likelihood which was taken to
be $2,000$, was used for each algorithm. For the MCMC algorithms,
$2,000$ MCMC iterations were used, with the first $500$ iterations
removed as burn in. In ABC-MCMC, the ABC tolerance was chosen to be
$\epsilon=0$ in both models, with $S_{1}\left(y\right)$ being the
statistic in the first order model and $\left(S_{1}\left(y\right),S_{2}\left(y\right)\right)$
the statistic in the second order model. SAV-MCMC required an estimate
$\hat{\theta}$ to construct the SAV estimator. For the first $500$
iterations this was taken to be 0 for the first order model and $\left(0,0\right)$
for the second order model, then for the remaining iterations this
was taken to be the sample average of the previous 250 iterations.
For the marginal SMC algorithms, 200 particles were used with $T=10$
target distributions. The sequence of target distributions used the
annealing scheme described in section \ref{subsec:Marginal-SMC-with}
with $\nu_{t}=\left(t/T\right)^{2}$. The results were compared with
the estimated posterior expectation obtained through a long run ($100,000$
iterations with $1,000$ iterations of burn in) of the exchange algorithm,
which was taken to be the ground truth. The ground truth is, to 5
s.f, for the first order model $\mathbb{E}_{\pi}\left[\theta_{1}\right]=0.27979$
and for the second order model $\mathbb{E}_{\pi}\left[\theta_{1}\right]=0.36134$
and $\mathbb{E}_{\pi}\left[\theta_{2}\right]=0.080639$.

\begin{table}
\begin{tabular}{|c|c|c|c|c|c|}
\hline 
Algorithm & Exchange & SAV-MCMC & ABC-MCMC & SAV-mSMC & Path mSMC\tabularnewline
\hline 
\hline 
Bias & $9.68\times10^{-4}$ & $7.96\times10^{-3}$ & \textcolor{red}{$-6.12\times10^{-2}$} & \textcolor{green}{$1.02\times10^{-5}$} & $9.62\times10^{-4}$\tabularnewline
\hline 
s.d. & $6.35\times10^{-3}$ & $1.32\times10^{-2}$ & \textcolor{red}{$9.21\times10^{-2}$} & $1.11\times10^{-2}$ & \textcolor{green}{$4.81\times10^{-3}$}\tabularnewline
\hline 
RMSE & $6.43\times10^{-3}$ & $1.32\times10^{-2}$ & \textcolor{red}{$1.14\times10^{-1}$} & $1.11\times10^{-2}$ & \textcolor{green}{$4.90\times10^{-3}$}\tabularnewline
\hline 
\end{tabular}

\caption{Estimates of bias, standard deviation and root mean square error for
estimates of the expectation of $\theta$ in the first order Ising
model (rounded to 3 s.f.).\label{tab:Estimates-of-bias,}}
\end{table}

\begin{table}
\begin{tabular}{|c|c|c|c|c|c|}
\hline 
Algorithm & Exchange & SAV-MCMC & ABC-MCMC & SAV-mSMC & Path-mSMC\tabularnewline
\hline 
\hline 
Bias ($\theta_{1}$) & \textcolor{green}{$4.17\times10^{-5}$} & $-1.41\times10^{-2}$ & \textcolor{red}{$3.42\times10^{-1}$} & $-1.58\times10^{-3}$ & $1.30\times10^{-3}$\tabularnewline
\hline 
s.d. ($\theta_{1}$) & $1.95\times10^{-2}$ & $3.72\times10^{-2}$ & \textcolor{red}{$5.71\times10^{-2}$} & $4.14\times10^{-2}$ & \textcolor{green}{$1.57\times10^{-2}$}\tabularnewline
\hline 
RMSE ($\theta_{1}$) & $1.95\times10^{-2}$ & $3.98\times10^{-2}$ & \textcolor{red}{$3.47\times10^{-1}$} & $4.14\times10^{-2}$ & \textcolor{green}{$1.57\times10^{-2}$}\tabularnewline
\hline 
Bias ($\theta_{2}$) & $8.00\times10^{-4}$ & $9.18\times10^{-3}$ & \textcolor{red}{$8.68\times10^{-2}$} & \textcolor{green}{$1.57\times10^{-4}$} & $1.24\times10^{-3}$\tabularnewline
\hline 
s.d. ($\theta_{2}$) & $1.61\times10^{-2}$ & \textcolor{red}{$3.41\times10^{-2}$} & $2.17\times10^{-2}$ & $2.83\times10^{-2}$ & \textcolor{green}{$1.26\times10^{-2}$}\tabularnewline
\hline 
RMSE ($\theta_{2}$) & $1.61\times10^{-2}$ & $3.53\times10^{-2}$ & \textcolor{red}{$8.94\times10^{-2}$} & $2.83\times10^{-2}$ & \textcolor{green}{$1.27\times10^{-2}$}\tabularnewline
\hline 
\end{tabular}

\caption{Estimates of bias, standard deviation and root mean square error for
estimates of the expectation of $\theta_{1}$ and $\theta_{2}$ in
the second order Ising model (rounded to 3 s.f.).\label{tab:Estimates-of-bias,-1}}
\end{table}

In addition to the methods here, several other algorithms were run.
Most notably, we considered a different way of reusing simulations
from the likelihood: using regression. In the spirit of papers such
as \citet{Sherlock2017}, instead of constructing a path estimate
using previous samples, we fit a linear regression model with the
response being the estimated log-likelihood and the covariates being
the values of $\theta$ at which the likelihood was estimated. We
found that this was not competitive with other approaches, since a
large bias was introduced through the estimate being performed in
$\log$-space. We also compared our SMC approach with the method of
\citet{Koblents2015} which, instead of using an annealed sequence
of targets, uses an annealing (or truncation) scheme to regularise
the weights of a PMC algorithm, with the annealing changing as the
algorithm progresses. We found that the performance of the two approaches
is comparable.

\begin{figure}
\subfloat[The ESS of the SMC sample at each SMC iteration before resampling
for a run of SAV-mSMC and path-mSMC.\label{fig:The-effective-sample}]{\includegraphics[scale=0.2]{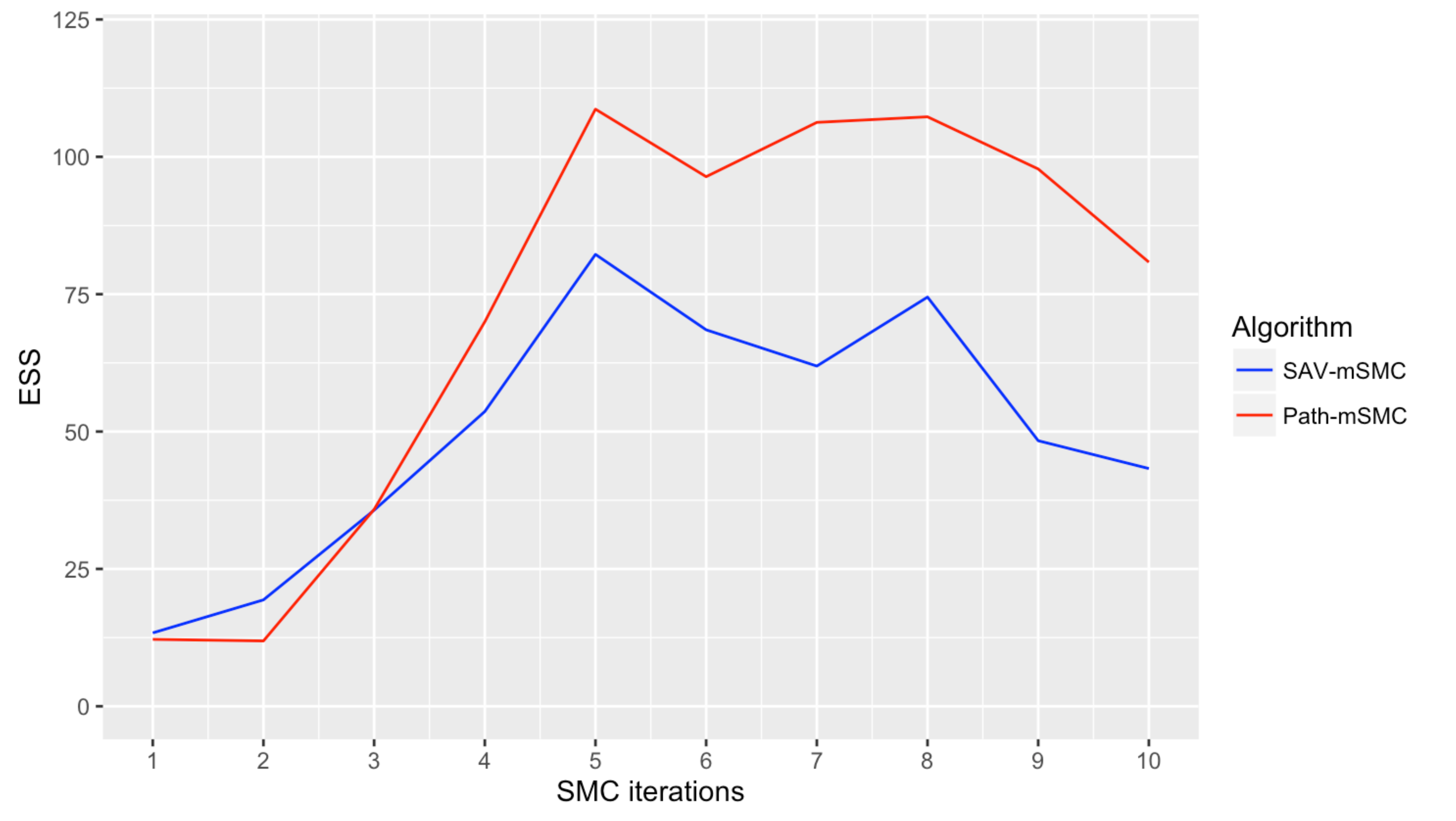}

}\subfloat[A sequence of samples after resampling from the path-mSMC algorithm
applied to the Ising model data. Points from the first annealed target
are shown in the lightest grey, with a darker grey for each successive
SMC iteration, finishing with black for the final target.\label{fig:A-sequence-of}]{\includegraphics[scale=0.2]{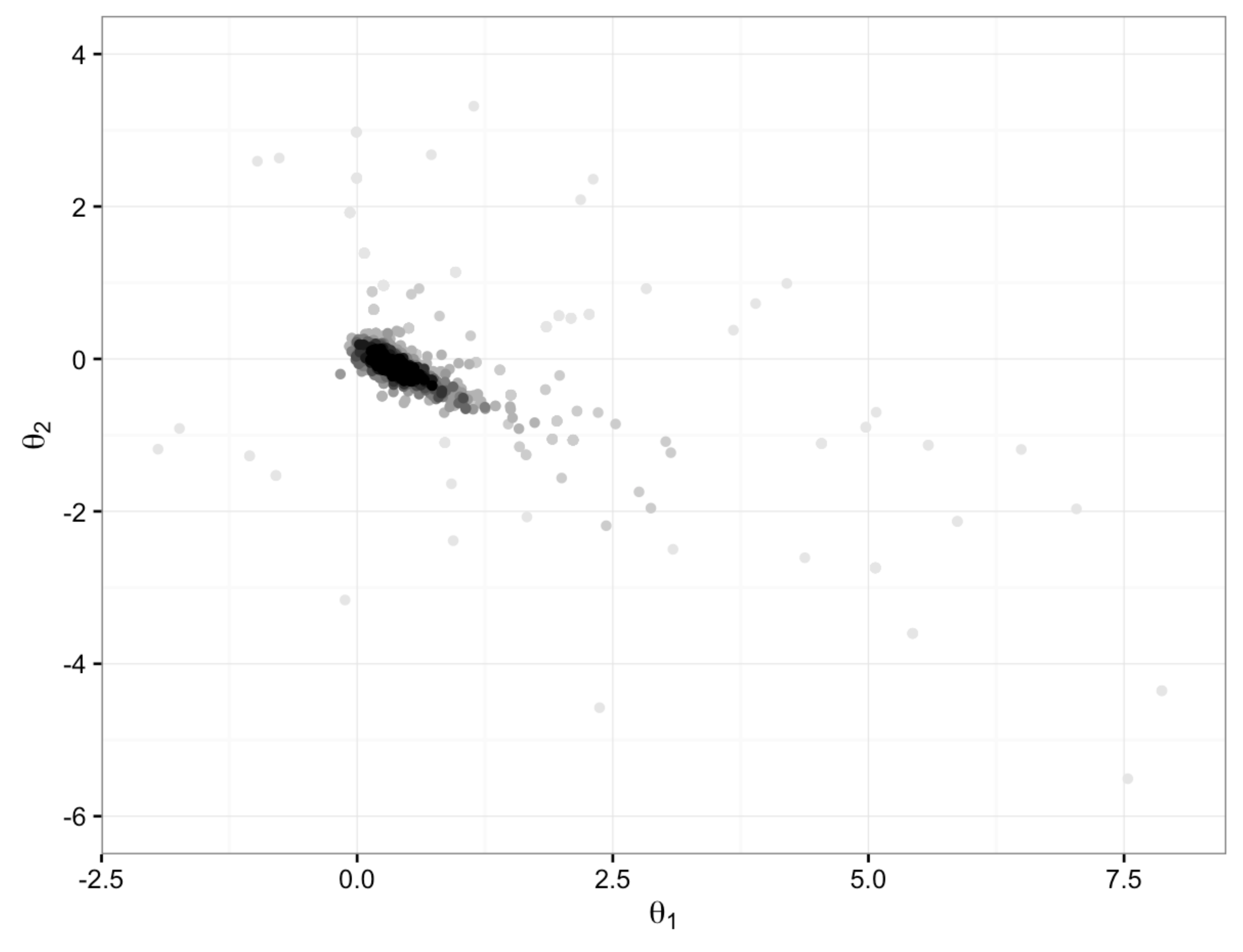}}

\caption{Results from marginal SMC algorithms on the Ising model data.}

\end{figure}

Tables \ref{tab:Estimates-of-bias,} and \ref{tab:Estimates-of-bias,-1}
give the bias, standard deviation and root mean square error for estimates
of the expectation from the different algorithms. We see that ABC-MCMC
exhibits the worst performance: the restriction that the ABC tolerance
is zero leads to high variance likelihood estimates, giving an inefficient
algorithm. As is emphasised in appendix \ref{sec:Connecting-ABC-with},
in ABC the use of the simulations from $x$-space is less efficient
than for the other methods. SAV-MCMC is outperformed by the exchange
algorithm: this is likely due to the fact that the exchange algorithm
directly estimates the ratio of partition functions, compared to SAV
which uses a ratio of estimates of the reciprocal of partition functions.
The results of SAV-mSMC are comparable to SAV-MCMC: there are fewer
Monte Carlo points in the SMC approach, but the effective size of
the MCMC sample is reduced by the autocorrelation in the chain. Path-mSMC
always outperforms SAV-mSMC due to the reduced variance of the likelihood
estimates. Figure \ref{fig:The-effective-sample} highlights the improvement
of path-mSMC though showing the improved effective sample size (ESS)
\citep{Kong1994} over SAV-mSMC. Figure \ref{fig:A-sequence-of},
which shows the SMC sample over the SMC iterations illustrates why
the improvement in ESS is enhanced in the later iterations: the SMC
sample concentrates in a small region of the parameter space after
around iteration 5, leading to more useful samples being available
for the path estimator. Path-mSMC also outperforms the exchange algorithm
on this example: in this case the autocorrelation in the sample from
the exchange algorithm is outweighed by reduction in variance in the
likelihood estimates in path-mSMC.

\subsection{Exponential random graph model}

Exponential random graph models (ERGMs) are widely used in the analysis
of social networks. They describe how distributions over possible
networks through weighting the influence of local statistics of the
network. The model takes the form
\[
f\left(y\mid\theta\right)=\exp\left(\theta^{T}S\left(y\right)\right)/Z\left(\theta\right),
\]
with $S\left(y\right)$ being some vector of statistics. Bayesian
inference has only relatively recently \citep{Caimo2011} been used
for these models, with the exchange algorithm being the most widely
used approach, as implemented in the \texttt{Bergm} R package \citep{Caimo2014}.

\begin{figure}
\subfloat[The Dolphin network.\label{fig:The-Dolphin-network.-1}]{\includegraphics[scale=0.2]{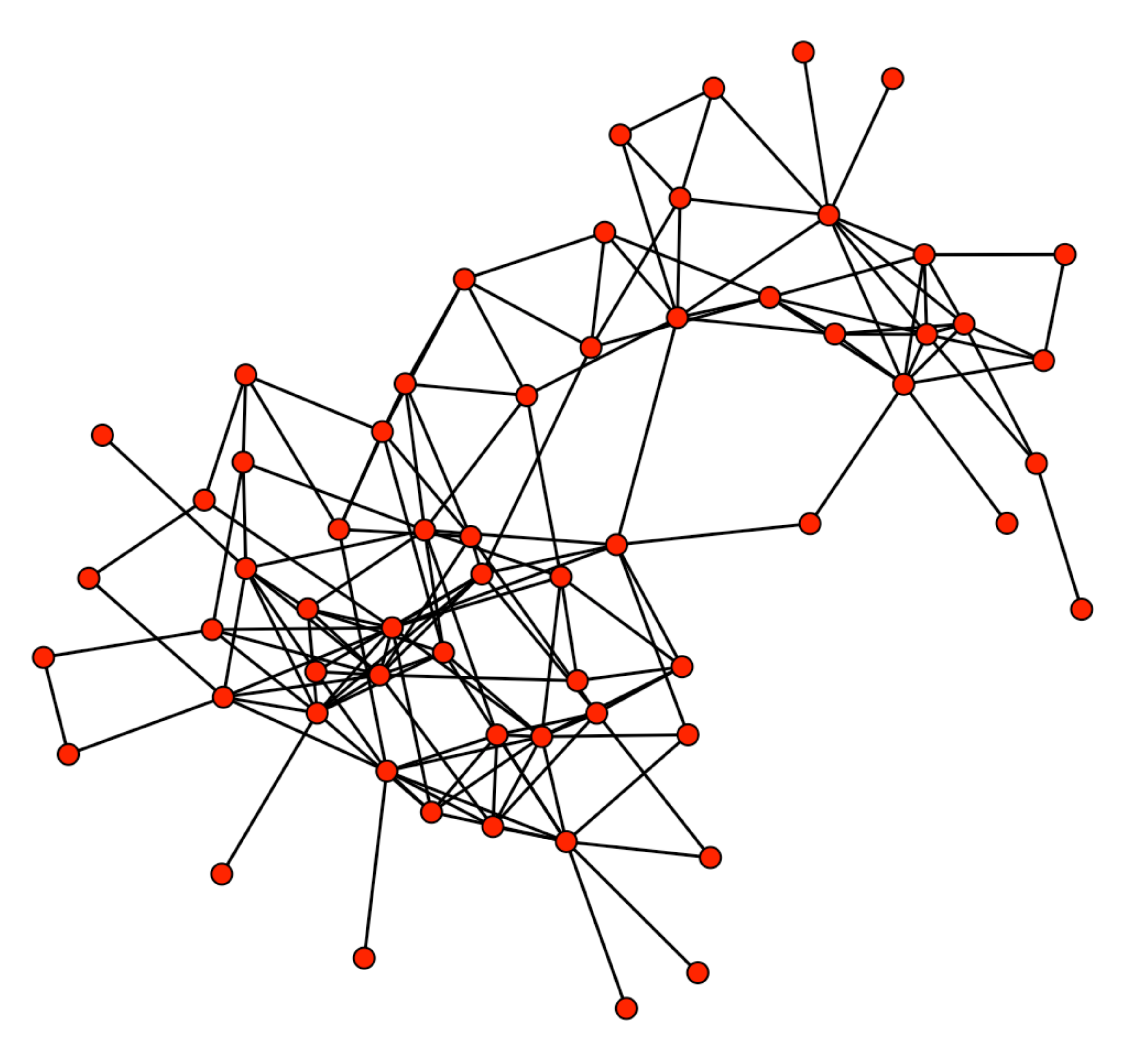}

}\subfloat[The ESS of the SMC sample at each SMC iteration before resampling
for a run of SAV-mSMC and path-mSMC.\label{fig:The-effective-sample-1}]{\includegraphics[scale=0.2]{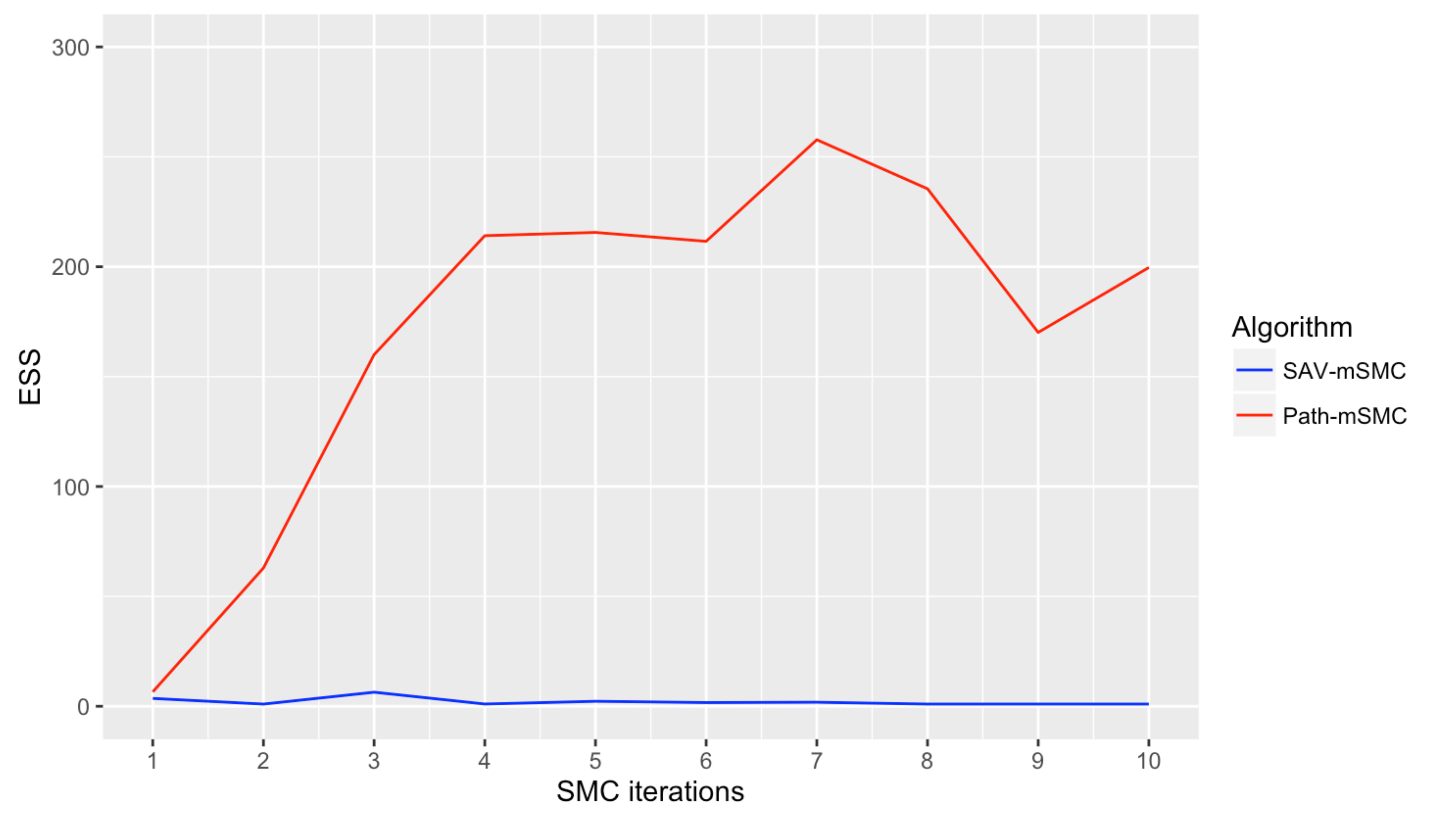}

}

\subfloat[A sequence of samples from the path-mSMC algorithm for $\theta_{1}$
and $\theta_{2}$.\label{fig:A-sequence-of-1}]{\includegraphics[scale=0.15]{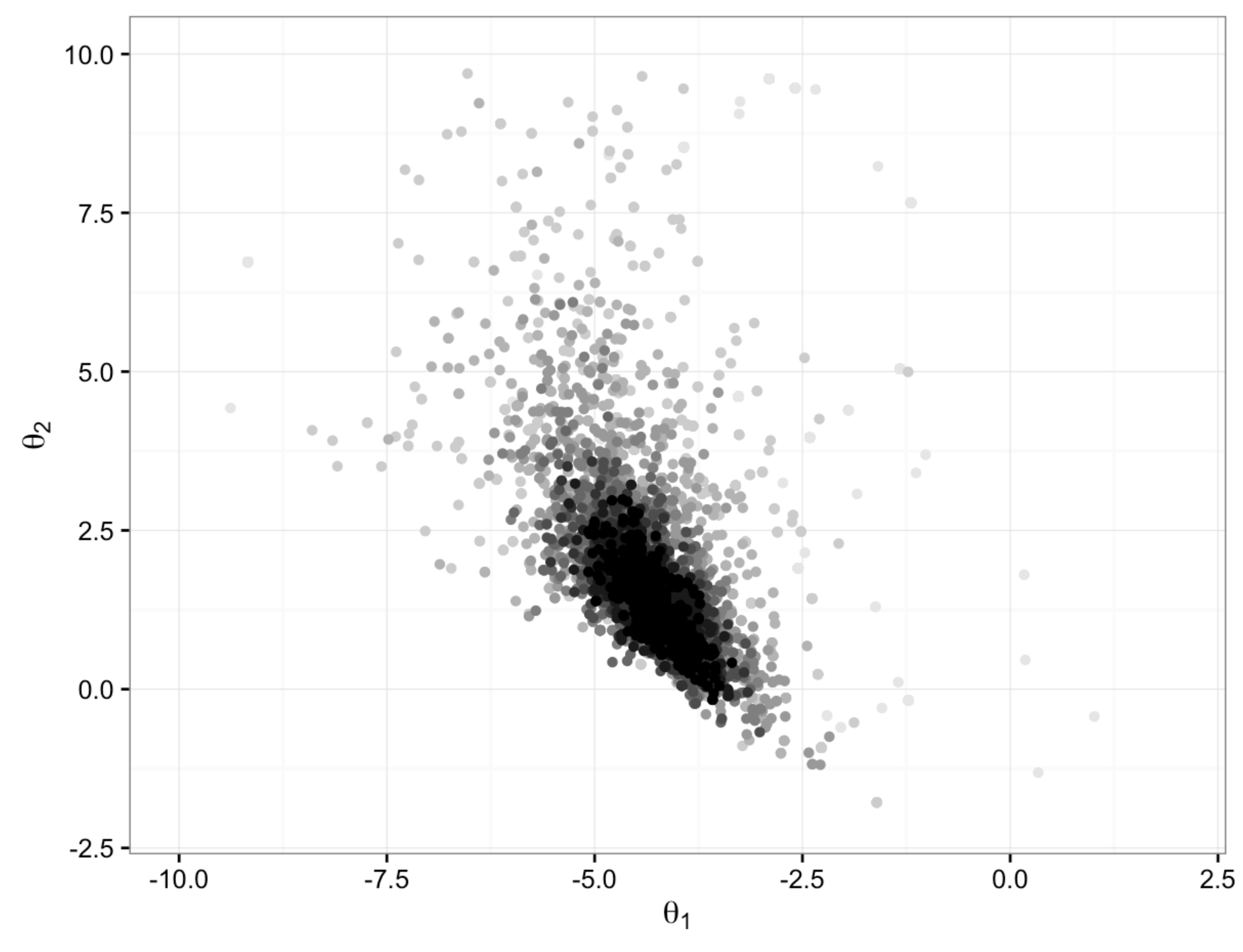}}\subfloat[A sequence of samples from the path-mSMC algorithm for $\theta_{2}$
and $\theta_{3}$.\label{fig:A-sequence-of-2}]{\includegraphics[scale=0.15]{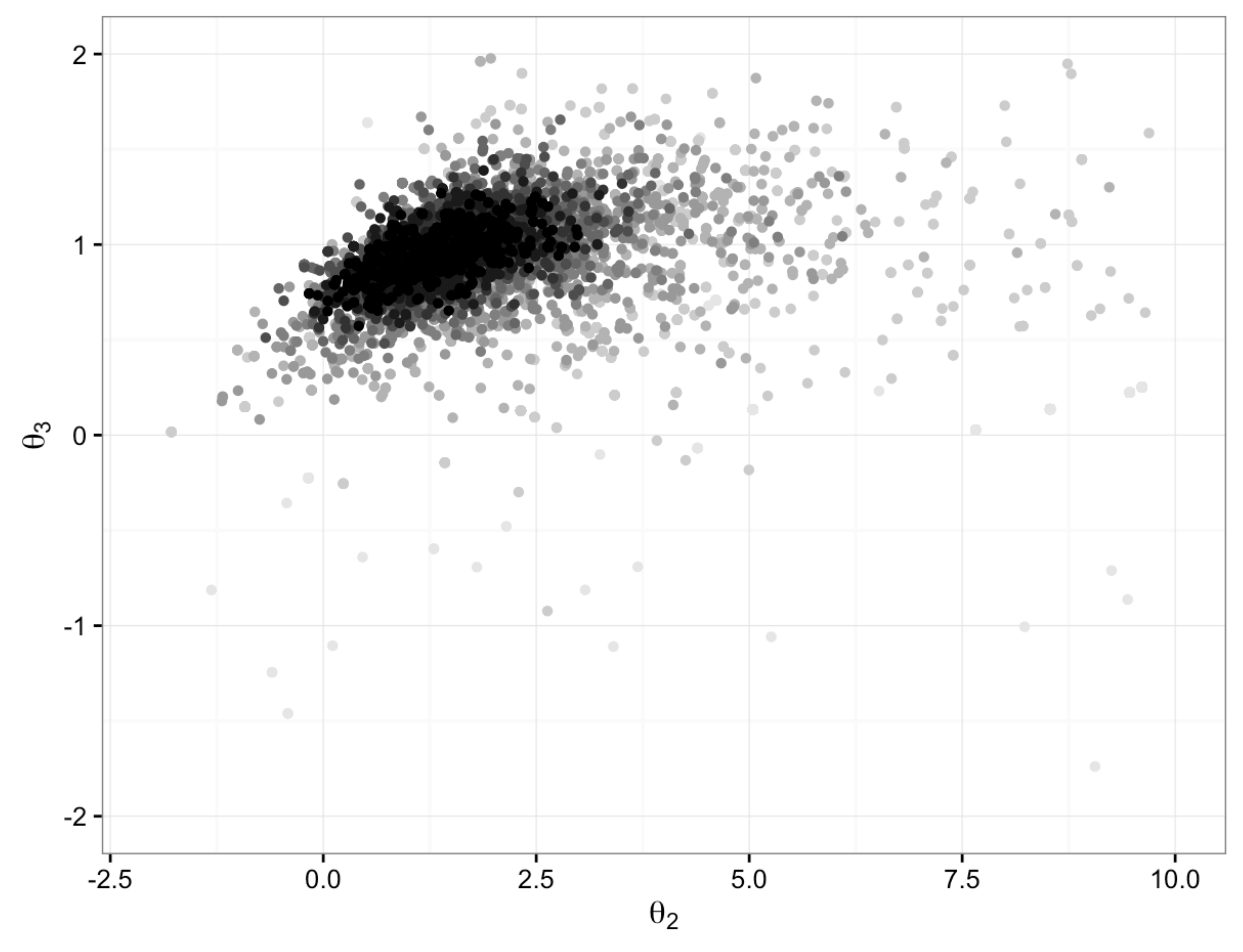}}\subfloat[A sequence of samples from the path-mSMC algorithm for $\theta_{1}$
and $\theta_{3}$.\label{fig:A-sequence-of-3}]{\includegraphics[scale=0.15]{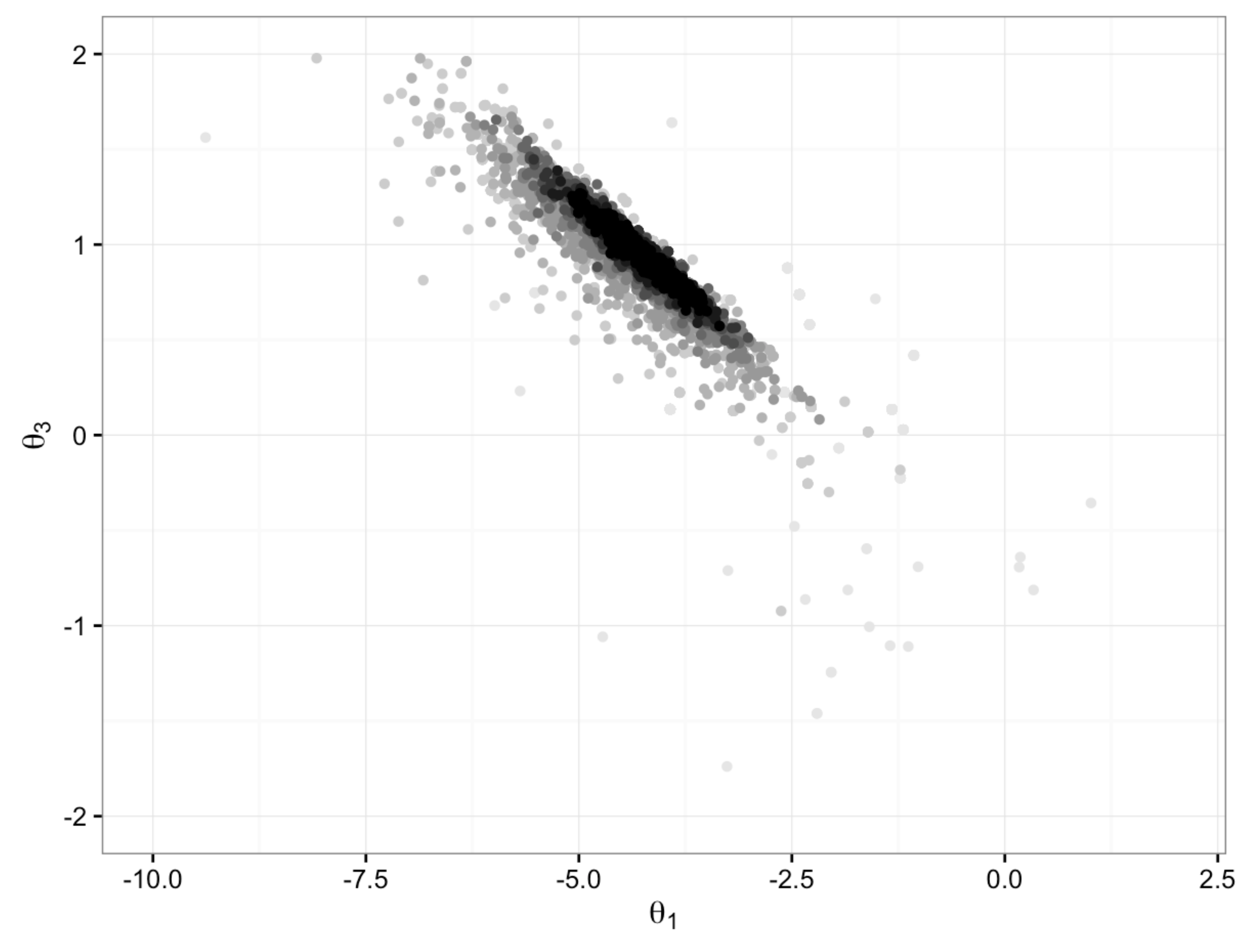}}

\caption{Results from the SMC algorithms applied to the Dolphin network.\label{fig:The-Dolphin-network.}
Where SMC samples are displayed, points from the first annealed target
are shown in the lightest grey, with a darker grey for each successive
SMC iteration, finishing with black for the final target.}
\end{figure}

In this section we compare the performance of the exchange algorithm
to the new path marginal SMC approach. We study the Dolphin network
\citep{Lusseau2003,Caimo2011}, shown in figure \ref{fig:The-Dolphin-network.-1},
and use the same statistics and priors as in \citet{Caimo2011}. The
statistics are
\begin{eqnarray*}
S_{1}(y)=\sum_{i<j}y_{ij} &  & \text{the number of edges}\\
S_{2}(y)=\exp\left(\phi_{u}\right)\sum_{i=1}^{n-1}\left\{ 1-\left(1-\exp\left(-\phi_{u}\right)\right)^{i}\right\} D_{i}\left(y\right) &  & \text{geometrically weighted degree}\\
S_{3}(y)=\exp\left(\phi_{v}\right)\sum_{i=1}^{n-2}\left\{ 1-\left(1-\exp\left(-\phi_{v}\right)\right)^{i}\right\} EP_{i}\left(y\right) &  & \text{geometrically weighted edgewise shared partner}
\end{eqnarray*}
and the prior \foreignlanguage{british}{on $\theta$ is $\theta\sim\mathcal{N}(0,30I_{3})$.
The \texttt{ergm} package \citep{Hunter2008} in R was used to simulate
from $f\left(\cdot\mid\theta\right)$, which uses the \textquotedblleft tie
no tie\textquotedblright{} (TNT) sampler. $15,000$ iterations of
the TNT sampler were used, and the $x$ sample was taken to be the
final one of these points.}

We ran both the exchange algorithm and path-mSMC 40 times, with both
algorithms having the same computational budget of $10,000$ simulations
from the likelihood. The exchange algorithm was run using the \texttt{Bergm}
software. We used two configurations of \texttt{Bergm}: the first
using a single chain for $10,000$ iterations with the first $1,000$
iterations discarded as burn in; the second using 6 parallel interacting
chains in the same configuration as recommended in \citet{Caimo2011},
with 1667 iterations per chain each with the first $1,000$ iterations
discarded as burn in (discarding fewer iterations made relatively
insignificant changes to the results). The SMC algorithm was run with
$1,000$ particles with $T=10$ SMC targets, again using the annealing
scheme with $\nu_{t}=\left(t/T\right)^{2}$. SAV-mSMC did not give
useful results for this model, due to the high variance of the estimates
of the reciprocal of the partition function. Figure \ref{fig:The-effective-sample-1}
shows the ESS of SAV-mSMC compared to path-mSMC: for many of the iterations
the ESS in SAV-mSMC is only marginally above one.

Table \ref{fig:The-effective-sample-1} gives the bias, standard deviation
and root mean square error for estimates of the expectation from the
exchange and path-mSMC algorithms, and sequences of samples from path-mSMC
are shown in figures \ref{fig:A-sequence-of-1}, \ref{fig:A-sequence-of-2}
and \ref{fig:A-sequence-of-3}. As in the previous section, path-mSMC
slightly outperforms the exchange algorithm in the main, for the same
reasons as in the previous application. For the expectation of $\theta_{2}$
(and also $\theta_{1}$ and $\theta_{3}$ to a lesser degree), the
variance of the path-mSMC estimate is increased by one of the 40 runs
(without including this run the estimated s.d.s for $\theta_{1}$.
$\theta_{2}$ and $\theta_{3}$ respectively are $3.72\times10^{-2}$,
$5.37\times10^{-2}$ and $1.47\times10^{-2}$). In this run the ESS
was so low in the first iteration that subsequent samples are affected,
and the algorithm does not recover. This issue may be rectified by
using additional target distributions in between the prior and the
first annealed target.

\begin{table}
\begin{tabular}{|c|c|c|c|}
\hline 
Algorithm & Exchange (single chain) & Exchange (6 chains) & Path-mSMC\tabularnewline
\hline 
\hline 
s.d. ($\theta_{1}$) & $5.58\times10^{-2}$ & ${\color{red}7.33\times10^{-2}}$ & \textcolor{green}{$5.52\times10^{-2}$}\tabularnewline
\hline 
s.d. ($\theta_{2}$) & ${\color{green}1.04\times10^{-1}}$ & $1.25\times10^{-1}$ & \texttt{\textcolor{red}{$4.00\times10^{-1}$}}\tabularnewline
\hline 
s.d. ($\theta_{3}$) & $1.97\times10^{-2}$ & \texttt{\textcolor{red}{$2.65\times10^{-2}$}} & ${\color{green}1.55\times10^{-2}}$\tabularnewline
\hline 
\end{tabular}

\caption{Estimates of bias, standard deviation and root mean square error for
estimates of the expectation of $\theta_{1}$, $\theta_{2}$ and $\theta_{3}$
in the ERGM (rounded to 3 s.f.).\label{tab:Estimates-of-bias,-2}}
\end{table}

\section{Discussion\label{sec:Discussion}}

This paper describes an SMC approach to Bayesian inference of parameters
of doubly intractable distributions, through investigating the use
of marginal SMC and showing how simulations from earlier iterations
of the SMC may be used to reduce the variance of likelihood estimates
at future iterations. This offers an alternative to the growing literature
on MCMC approaches based on the SAV method and the exchange algorithm
\citep{Moller2006,Murray2006}, and the IS and previous SMC approach
in \citet{Everitt2017}. Just as in any other situation (outside of
the doubly intractable setting), there are cases in which SMC has
advantages over MCMC: primarily that with MCMC it can be more difficult
to adapt proposal distributions and more difficult to estimate marginal
likelihoods. Further, SMC offers a natural way to use a population
of Monte Carlo points which may be useful for multi-modal posteriors.
In addition, as observed previously in the case of ABC \citep{Sisson2007f},
SMC may offer advantages over MCMC in the case where estimated likelihoods
are used.

In most settings it is more efficient to use an SMC sampler with MCMC
moves compared to the marginal SMC approach (see \citet{Didelot2011}
for a practical example of this). However, in the case of doubly intractable
distributions we observe two advantages of marginal SMC: firstly there
is the flexibility to use any sequence of distributions, where the
SMC sampler with MCMC moves is restricted to the case of data point
tempered distributions; secondly there is a natural way of reusing
previous simulations from the likelihood. We demonstrate good empirical
performance for the path-mSMC algorithm, which exploits both of these
advantages, and show that without reusing previous simulations the
performance of the marginal SMC algorithm may be poor.

Another potential advantage to the approach in this paper is that
all of the SMC targets but the final one may be approximate, which
may be advantageous if computationally cheap approximate posterior
distributions are available. In applications similar to those in this
paper, \citet{Everitt2017b} make use of approximate targets that
use shorter MCMC runs for simulating from the likelihood. We may use
such approximations at the early stages of marginal SMC, although
it may be difficult to automatically choose which approximation to
use at each SMC iteration.

Despite these advantages, there are some limitations to the use of
path-mSMC. Most importantly, we expect it to only be effective in
low to moderate dimensions: in part due to the use of marginal SMC,
and in part due to the reduced effectiveness of the path estimator
as the dimension increases.

\appendix

\section{Connecting ABC with the multiple auxiliary variable method\label{sec:Connecting-ABC-with}}

\subsection{ABC for doubly intractable models\label{subsec:ABC-for-doubly}}

Suppose that the model $f(y|\theta)$ has an intractable likelihood
but can be targeted by a MCMC chain $x=(x_{1},x_{2},\ldots,x_{n})$.
Let $\pi$ represent densities relating to this chain. Then $\pi_{n}(y|\theta):=\pi(x_{n}=y|\theta)$
is an approximation of $f(y|\theta)$ which can be estimated by ABC.
For now suppose that $y$ is discrete and consider the ABC likelihood
estimate requiring an exact match: simulate from $\pi(\mathbf{x}|\theta)$
and return $\mathbbm{1}(x_{n}=y)$. We will consider an IS variation
on this: simulate from $g(x|\theta)$ and return $\mathbbm{1}(x_{n}=y)\pi(x|\theta)/g(x|\theta)$.
Under the mild assumption that $g(x|\theta)$ has the same support
as $\pi(x|\theta)$ (typically true unless $n$ is small), both estimates
have the expectation $\Pr(x_{n}=y|\theta)$.

This can be generalised to cover continuous data using the identity
\[
\pi_{n}(y|\theta)=\int_{x_{n}=y}\pi(\mathbf{x}|\theta)dx_{1:n-1},
\]
 where $x_{i:j}$ represents $(x_{i},x_{i+1},\ldots,x_{j})$. An importance
sampling estimate of this integral is 
\begin{equation}
w=\frac{\pi(\mathbf{x}|\theta)}{g(x_{1:n-1}|\theta)}\label{eq:ISestimate}
\end{equation}
where $\mathbf{x}$ is sampled from $g(x_{1:n-1}|\theta)\delta(x_{n}=y)$,
with $\delta$ representing a Dirac delta measure. Then, under mild
conditions on the support of $g$, $w$ is an unbiased estimate of
$\pi_{n}(y|\theta)$.

The ideal choice of $g(x_{1:n-1}|\theta)$ is$\pi(x_{1:n-1}|x_{n},\theta)$,
as then $w=\pi(x_{n}=y|\theta)$ exactly. This represents sampling
from the Markov chain conditional on its final state being $y$.

\subsection{Equivalence to MAV\label{subsec:Equivalence-to-MAV}}

We now show that natural choices of $\pi(\mathbf{x}|\theta)$ and
$g(x_{1:n-1}|\theta)$ in the ABC method just outlined results in
the MAV estimator \ref{eq:mav}. Our choices are 
\begin{align*}
g(x_{1:n-1}|\theta) & =\prod_{i=1}^{n-1}K_{i}(x_{i}|x_{i+1})\\
\pi(\mathbf{x}|\theta) & =f_{1}(x_{1}|\theta,y)\prod_{i=1}^{n-1}K_{i}(x_{i+1}|x_{i}).
\end{align*}
Here $\pi(\mathbf{x}|\theta)$ defines a MCMC chain with transitions
$K_{i}(x_{i+1}|x_{i})$. Suppose $K_{i}$ is a reversible Markov kernel
with invariant distribution $f_{i}$ for $i\leq a$, and for $i>a$
it is a reversible Markov kernel with invariant distribution $f(\cdot|\theta)$.
Also assume $b:=n-a\to\infty$. Then the MCMC chain ends in a long
sequence of steps targeting $f(\cdot|\theta)$ so that $\lim_{n\to\infty}\pi_{n}(\cdot|\theta)=f(\cdot|\theta)$.
Thus the likelihood being estimated converges on the true likelihood
for large $n$. Note this is the case even for fixed $a$.

The importance density $g(x_{1:n-1}|\theta)$ specifies a reverse
time MCMC chain starting from $x_{n}=y$ with transitions $K_{i}(x_{i}|x_{i+1})$.
Simulating $\mathbf{x}$ is straightforward by sampling $x_{n-1}$,
then $x_{n-2}$ and so on. This importance density is an approximation
to the ideal choice stated at the end of section \ref{subsec:ABC-for-doubly}.

The resulting likelihood estimator is 
\[
w=f_{1}(x_{1}|\theta,y)\prod_{i=1}^{n-1}\frac{K_{i}(x_{i+1}|x_{i})}{K_{i}(x_{i}|x_{i+1})}.
\]
Using detailed balance gives 
\[
\frac{K_{i}(x_{i+1}|x_{i})}{K_{i}(x_{i}|x_{i+1})}=\frac{f_{i}(x_{i+1}|\theta,y)}{f_{i}(x_{i}|\theta,y)}=\frac{\gamma_{i}(x_{i+1}|\theta,y)}{\gamma_{i}(x_{i}|\theta,y)},
\]
so that 
\[
w=f_{1}(x_{1}|\theta,y)\prod_{i=1}^{n-1}\frac{\gamma_{i}(x_{i+1}|\theta,y)}{\gamma_{i}(x_{i}|\theta,y)}=\gamma(y|\theta)\prod_{i=2}^{n}\frac{\gamma_{i-1}(x_{i}|\theta,y)}{\gamma_{i}(x_{i}|\theta,y)}.
\]
This is an unbiased estimator of $\pi_{n}(y|\theta)$. Hence 
\[
v=\prod_{i=2}^{n}\frac{\gamma_{i-1}(x_{i}|\theta,y)}{\gamma_{i}(x_{i}|\theta,y)}=\prod_{i=2}^{a}\frac{\gamma_{i-1}(x_{i}|\theta,y)}{\gamma_{i}(x_{i}|\theta,y)}.
\]
is an unbiased estimator of $\pi_{n}(y|\theta)/\gamma(y|\theta)\to1/Z(\theta)$.
 In the above we have assumed, as in section \ref{subsec:ABC-for-doubly},
that $\gamma_{1}$ is normalised. When this is not the case then we
instead get an estimator of $Z(\hat{\theta})/Z(\theta)$, as for MAV
methods. Also note that in either case a valid estimator is produced
for any choice of $y$.

We note that this carefully designed ABC estimate is precisely the
same as the MAV approach. We may view the method as a two stage procedure.
First run a MCMC chain of length $b$ with any starting value, targeting
$f(\cdot|\theta)$. Let its final value be $x_{a}$. Secondly run
a MCMC chain $x_{a},x_{a-1},\ldots$ using kernels $K_{a-1},K_{a-2},\ldots$
and evaluate the estimator $v$. This is unbiased in the limit $b\to\infty$,
so the first stage could be replaced by perfect sampling methods where
these exist.

\subsection{Remark}

In this section we have illustrated that a carefully constructed ABC
approach (in which the full data is used instead of a summary) yields
the same algorithm as the MAV method. We might MAV method to result
in an improvement over the standard ABC algorithm due to the more
effective use of the simulations from the model $f$. In fact, in
section \ref{sec:Empirical-results} we observe empirically that even
the highest variance version of MAV (i.e. the original SAV approach)
outperforms ABC in which sufficient statistics are used, i.e. the
lowest variance version of ABC.

\bibliographystyle{/Users/Everitt/Dropbox/projects/bib/mychicago}
\bibliography{/Users/Everitt/Dropbox/projects/abc_fewer_data/z_regression/z_regression}

\end{document}